\documentclass{article}
\usepackage{arxiv}
\usepackage[utf8]{inputenc}
\usepackage[T1]{fontenc}
\usepackage{hyperref}
\usepackage{url}
\usepackage{booktabs}
\usepackage{amsfonts}
\usepackage{amsmath}
\usepackage{amssymb}
\usepackage{nicefrac}
\usepackage{microtype}
\usepackage{cleveref}
\usepackage{graphicx}
\usepackage{natbib}
\usepackage{doi}
\usepackage{bm}
\usepackage{multirow}
\usepackage{subcaption}

\graphicspath{{figures/}}
\title{Model Reference Adaptive Control for Gust Load Alleviation of Nonlinear Aeroelastic Systems}

\author{
  Nikolaos D.~Tantaroudas\thanks{Corresponding author. Senior Researcher, ICCS.} \\
  Institute of Communications and Computer Systems (ICCS)\\
  9 Iroon Politechniou Street, Zografou, Athens 15773, Greece \\
  \texttt{nikolaos.tantaroudas@iccs.gr} \\
  \And
  Guanqun Gai \\
  School of Engineering, University of Liverpool\\
  Liverpool L69 3BX, United Kingdom \\
  \And
  Ilias Karachalios \\
  National Technical University of Athens\\
  Zografou, Athens 15780, Greece \\
}

\renewcommand{\shorttitle}{MRAC for Gust Load Alleviation}

\hypersetup{
  pdftitle={Model Reference Adaptive Control for Gust Load Alleviation of Nonlinear Aeroelastic Systems},
  pdfsubject={eess.SY, physics.flu-dyn},
  pdfauthor={N.D. Tantaroudas, G. Gai, I. Karachalios},
  pdfkeywords={Adaptive control, MRAC, Gust load alleviation, Lyapunov stability, Flexible aircraft}
}

\begin{document}
\maketitle

\begin{abstract}
Model Reference Adaptive Control (MRAC) based on Lyapunov stability theory is developed for gust load alleviation (GLA) of nonlinear aeroelastic systems. The controller operates on a nonlinear reduced-order model (NROM) derived from Taylor series expansion and eigenvector projection of the coupled fluid--structure--flight dynamic equations. The complete MRAC formulation is presented, including the reference model design that encodes desired closed-loop damping characteristics, the adaptive control law with real-time gain adjustment, and the Lyapunov-based derivation of the adaptation law that guarantees asymptotic tracking in the linear case and bounded tracking under a Lipschitz condition on the nonlinear residual. The adaptation rate matrix is identified as the single most important design parameter, governing the trade-off between convergence speed, peak load reduction, and actuator demand. Two test cases are considered: (i)~a three-degree-of-freedom aerofoil with cubic stiffness nonlinearities, and (ii)~a Global Hawk-like unmanned aerial vehicle (UAV) with 540 structural degrees of freedom. For the UAV under a discrete ``1-cosine'' gust, MRAC achieves significant wing-tip deflection reductions, outperforming the $\mathcal{H}_\infty$ robust control benchmark with comparable control effort. Under Von K\'{a}rm\'{a}n stochastic turbulence, meaningful reductions are also obtained, with performance scaling with the adaptation rate. The results demonstrate that MRAC provides an effective framework for GLA of flexible aircraft operating in both deterministic and stochastic disturbance environments.
\end{abstract}


%% ============================================================
\section{Introduction}
\label{sec:intro}
%% ============================================================

\subsection{Motivation}
\label{sec:intro_motivation}

The trend towards lighter, more efficient airframe designs has led to increasingly flexible aircraft structures. High-altitude long-endurance (HALE) platforms, such as the Global Hawk and Helios, exhibit large wing-tip deflections under normal flight conditions~\citep{Patil2001, Noll2004, Patil2006}. At such deformation levels, geometric nonlinearities become significant and the coupling between structural dynamics, unsteady aerodynamics, and rigid-body flight mechanics can no longer be neglected~\citep{Cesnik2012, Hesse2014}. The Helios mishap in 2003, in which turbulence-induced structural deformations led to an unrecoverable divergence, underscored the critical importance of gust load alleviation (GLA) for very flexible aircraft~\citep{Noll2004}.

Active GLA systems employ control surfaces, typically trailing-edge flaps distributed along the wing span, to counteract the aerodynamic loads induced by atmospheric disturbances. The design of such systems requires a control strategy that can accommodate: (i)~significant model uncertainty, since the aeroelastic characteristics change with flight condition, payload, and fuel state; (ii)~time-varying plant dynamics arising from the nonlinear coupling; and (iii)~both deterministic gusts (as specified in certification requirements) and stochastic turbulence encountered in operational service~\citep{Livne2018, Dowell2004}.

\subsection{Robust versus Adaptive Control for GLA}
\label{sec:intro_robust_adaptive}

Robust control methods, particularly $\mathcal{H}_\infty$ synthesis, have been widely applied to aeroelastic systems and offer guaranteed worst-case performance bounds~\citep{Tantaroudas2015scitech, Badcock2011}. However, these methods rely on a fixed nominal plant model and must accommodate all anticipated uncertainties within a single controller design. This conservatism can limit the achievable GLA performance: the controller must be designed for the worst-case plant, which may differ substantially from the actual plant at any given operating point.

Adaptive control offers a fundamentally different paradigm. Rather than designing a single fixed controller for all anticipated conditions, the controller gains adjust in real time to match the actual plant dynamics. Model Reference Adaptive Control (MRAC) is a particularly attractive form of adaptive control for GLA because it drives the plant response to track a user-specified reference model~\citep{Tantaroudas2014aviation, Tantaroudas2017bookchapter}. The reference model can be designed to provide the desired level of structural damping and load alleviation, and the adaptive law ensures that the actual plant response converges to this ideal behaviour regardless of the initial model uncertainty.

\subsection{Literature Review}
\label{sec:intro_literature}

The application of adaptive control to aeroelastic systems has a growing literature. Su and Cesnik~\citep{SuCesnik2010} applied $L_1$ adaptive control to a very flexible blended-wing-body aircraft, demonstrating improved gust response compared to fixed-gain controllers. Da~Ronch et al.~\citep{DaRonch2014flutter} developed a nonlinear controller for flutter suppression using derivative-free methods, validating the approach from numerical simulation through to wind tunnel testing. Papatheou et al.~\citep{Papatheou2013ifasd} conducted experimental investigations of active flutter suppression using adaptive control on a pitch--plunge aerofoil rig. Fichera et al.~\citep{Fichera2014isma} combined experimental and numerical studies of nonlinear dynamic behaviour, providing validation data for control-oriented models. More recently, Wang et al.~\citep{Wang2018mpc} proposed a nonlinear aeroelastic control strategy based on model updating and model-predictive control for very flexible aircraft, while Wang et al.~\citep{Wang2019indi} developed an incremental nonlinear dynamic inversion approach for flexible aircraft GLA that simultaneously regulates rigid-body motions and suppresses elastic modes. Artola et al.~\citep{Artola2021} demonstrated aeroelastic control and state estimation using a minimal nonlinear modal description combined with moving-horizon estimation and model-predictive control. Riso and Cesnik~\citep{Riso2023} conducted a systematic assessment of low-order modelling approaches for aeroelastic predictions of very flexible wings. A comprehensive treatment of the coupled flight mechanics, aeroelasticity, and control of flexible aircraft has been provided in the monograph by Palacios and Cesnik~\citep{PalaciosCesnik2023}.

In the context of model reduction for control, Da~Ronch et al.~\citep{DaRonch2013control, DaRonch2013gust} developed techniques for reducing nonlinear aeroelastic models to forms suitable for real-time control implementation, and Da~Ronch et al.~\citep{DaRonch2012rom} demonstrated nonlinear model reduction for flexible aircraft control design. These reduced-order models (ROMs) provide the state-space representations on which MRAC can be designed.

The present work, based on the authors' earlier conference publications~\citep{Tantaroudas2014aviation, DaRonch2014scitech_flight} and the comprehensive treatment in~\citet{Tantaroudas2017bookchapter}, extends the MRAC framework to coupled aeroelastic--flight dynamic systems with nonlinear reduced-order models, provides a complete stability proof for the nonlinear case, and presents a systematic study of adaptation rate effects on both discrete gust and stochastic turbulence responses. The NMOR formulation underpinning the present controller is derived in detail in the companion paper~\citep{Tantaroudas2026nmor}, its application to rapid worst-case gust identification is demonstrated in~\citep{Tantaroudas2026gust}, and the $\mathcal{H}_\infty$ robust control counterpart used for comparison herein is presented in~\citep{Tantaroudas2026hinf}.

\subsection{Contributions and Paper Outline}
\label{sec:intro_contributions}

The specific contributions of this paper are as follows. First, a complete MRAC formulation for nonlinear aeroelastic systems is presented, including the reference model design, adaptive control law, error dynamics, Lyapunov stability proof, and adaptation law derivation. Second, the stability analysis is extended to the nonlinear case, establishing a Lipschitz condition on the nonlinear residual that guarantees bounded tracking error. Third, a systematic study of the adaptation rate matrix $\boldsymbol{\Gamma}$ and its effect on convergence speed, peak load reduction, and actuator demands is conducted. Fourth, the framework is validated on two test cases of increasing complexity: a three-degree-of-freedom aerofoil with cubic nonlinearities and a full-aircraft Global Hawk-like UAV. Finally, a comparison with $\mathcal{H}_\infty$ robust control is provided, demonstrating MRAC superiority for discrete gusts with comparable or reduced control effort.

The remainder of the paper is organised as follows. \Cref{sec:model} presents the aeroelastic modelling framework and the nonlinear model order reduction technique. \Cref{sec:mrac} develops the MRAC theory in detail, including the Lyapunov stability proof and the treatment of non-minimum phase dynamics. \Cref{sec:3dof} presents results for the three-degree-of-freedom aerofoil. \Cref{sec:uav} presents results for the Global Hawk-like UAV under discrete gust and stochastic turbulence. \Cref{sec:discussion} provides a discussion of adaptation rate effects and comparison with robust control. \Cref{sec:conclusions} draws conclusions.

%% ============================================================
\section{Aeroelastic Modelling and Reduction}
\label{sec:model}
%% ============================================================

\subsection{Coupled Fluid--Structure--Flight Dynamic System}
\label{sec:model_coupled}

The coupled aeroelastic--flight dynamic system is described by the nonlinear ordinary differential equations arising from the spatial discretisation of the structural (finite-element), aerodynamic (unsteady panel or vortex-lattice), and flight dynamic (rigid-body) sub-systems~\citep{Hesse2014, Murua2012, Palacios2010}. The structural model employs a geometrically exact beam formulation~\citep{Hodges2003, Palacios2010intrinsic} that captures the large deformations characteristic of very flexible wings. The unsteady aerodynamics are modelled using either a strip-theory formulation with Wagner and K\"{u}ssner indicial functions~\citep{Wagner1925, Kussner1936, Theodorsen1935} for the two-dimensional aerofoil case, or an unsteady vortex-lattice method~\citep{Murua2012} for the three-dimensional UAV case.

After spatial discretisation and linearisation about a nonlinear equilibrium, the system can be written in first-order form as~\citep{Tantaroudas2017bookchapter, Hesse2014}:
\begin{equation}
    \dot{\mathbf{w}} = \mathbf{A}_f\,\mathbf{w} + \mathbf{B}_{c,f}\,\mathbf{u}_c + \mathbf{B}_{g,f}\,\mathbf{u}_d + \mathbf{F}_{NL}(\mathbf{w})
    \label{eq:full_system}
\end{equation}
where $\mathbf{w} \in \mathbb{R}^{N}$ is the full-order state vector (with $N$ typically of the order of hundreds to thousands), $\mathbf{A}_f \in \mathbb{R}^{N \times N}$ is the Jacobian matrix evaluated at the equilibrium, $\mathbf{B}_{c,f} \in \mathbb{R}^{N \times m}$ is the control input matrix mapping the $m$ control surface deflections $\mathbf{u}_c$ to the state dynamics, $\mathbf{B}_{g,f} \in \mathbb{R}^{N \times p}$ is the gust input matrix mapping the $p$ gust disturbance inputs $\mathbf{u}_d$ to the state dynamics, and $\mathbf{F}_{NL}(\mathbf{w})$ contains the nonlinear terms arising from the Taylor series expansion of the full nonlinear residual about the equilibrium~\citep{DaRonch2013control}.

The nonlinear terms $\mathbf{F}_{NL}$ include second-order (quadratic) and third-order (cubic) contributions:
\begin{equation}
    \mathbf{F}_{NL}(\mathbf{w}) = \mathbf{F}_2(\mathbf{w},\mathbf{w}) + \mathbf{F}_3(\mathbf{w},\mathbf{w},\mathbf{w}) + \mathcal{O}(\|\mathbf{w}\|^4)
    \label{eq:nonlinear_terms}
\end{equation}
where $\mathbf{F}_2$ and $\mathbf{F}_3$ are multilinear operators representing the quadratic and cubic nonlinearities, respectively. For structural nonlinearities of the hardening type, the cubic terms dominate and tend to increase the effective stiffness at large deformations~\citep{Alighanbari1995, Irani2011}.

\subsection{Nonlinear Model Order Reduction (NMOR)}
\label{sec:model_nmor}

The full-order system~\eqref{eq:full_system} is too large for direct use in real-time adaptive control. A model order reduction is performed by projecting the state vector onto a small basis of eigenvectors~\citep{DaRonch2013control, DaRonch2013gust, Lucia2004}. Let $\boldsymbol{\Phi} \in \mathbb{R}^{N \times n}$ be the matrix of the $n \ll N$ dominant right eigenvectors of $\mathbf{A}_f$, and let $\boldsymbol{\Psi} \in \mathbb{R}^{n \times N}$ be the corresponding matrix of left eigenvectors such that $\boldsymbol{\Psi}\boldsymbol{\Phi} = \mathbf{I}_n$. The state is approximated as $\mathbf{w} \approx \boldsymbol{\Phi}\,\mathbf{x}$, where $\mathbf{x} \in \mathbb{R}^n$ is the reduced state vector. Substituting into~\eqref{eq:full_system} and premultiplying by $\boldsymbol{\Psi}$ yields the nonlinear reduced-order model (NROM):
\begin{equation}
    \dot{\mathbf{x}} = \mathbf{A}\,\mathbf{x} + \mathbf{B}_{c}\,\mathbf{u}_c + \mathbf{B}_{g}\,\mathbf{u}_d + \mathbf{F}_{NR}(\mathbf{x})
    \label{eq:nrom}
\end{equation}
where $\mathbf{A} = \boldsymbol{\Psi}\mathbf{A}_f\boldsymbol{\Phi} = \mathrm{diag}(\lambda_1, \ldots, \lambda_n)$ is the diagonal matrix of retained eigenvalues, $\mathbf{B}_{c} = \boldsymbol{\Psi}\mathbf{B}_{c,f}$, $\mathbf{B}_{g} = \boldsymbol{\Psi}\mathbf{B}_{g,f}$, and $\mathbf{F}_{NR}(\mathbf{x}) = \boldsymbol{\Psi}\mathbf{F}_{NL}(\boldsymbol{\Phi}\mathbf{x})$ is the projected nonlinear function.

The linear part of the ROM (obtained by setting $\mathbf{F}_{NR} = \mathbf{0}$) serves as the basis for reference model design and stability analysis, while the full NROM is used for plant simulation and to assess the effect of nonlinearities on the adaptive controller performance.

\subsection{Selection of Retained Modes}
\label{sec:model_modes}

The number and choice of retained eigenvectors is critical for the accuracy of the ROM. The selection is guided by two criteria: (i)~the modes must capture the dominant dynamic response to gust disturbances, and (ii)~the modes must include those that are most affected by the control surface inputs. In practice, the retained modes are the lowest-frequency aeroelastic modes that have significant participation in the gust response, as determined by the magnitude of the corresponding columns of $\mathbf{B}_g$. For the three-degree-of-freedom aerofoil, $n = 8$ reduced states are retained (three complex-conjugate eigenvalue pairs corresponding to the structural degrees of freedom and two real eigenvalues related to the gust influence). For the Global Hawk-like UAV, eight modes are retained from the full 540-DOF system (five complex-conjugate structural mode pairs and three real gust modes).

%% ============================================================
\section{Model Reference Adaptive Control}
\label{sec:mrac}
%% ============================================================

This section presents the MRAC formulation in detail, following the treatment in~\citet{Tantaroudas2017bookchapter} and~\citet{Tantaroudas2014aviation}. The development proceeds from the plant and reference model definitions through the adaptive control law, error dynamics, Lyapunov stability proof, and adaptation law derivation, concluding with the treatment of non-minimum phase dynamics.

\subsection{Plant Model}
\label{sec:mrac_plant}

The plant to be controlled is the NROM of~\cref{eq:nrom}:
\begin{equation}
    \dot{\mathbf{x}}(t) = \mathbf{A}\,\mathbf{x}(t) + \mathbf{B}_{c}\,\mathbf{u}_c(t) + \mathbf{B}_{g}\,\mathbf{u}_d(t) + \mathbf{F}_{NR}(\mathbf{x}(t))
    \label{eq:plant}
\end{equation}
where $\mathbf{x}(t) \in \mathbb{R}^n$ is the plant state, $\mathbf{u}_c(t) \in \mathbb{R}^m$ is the control input (flap deflections), $\mathbf{u}_d(t) \in \mathbb{R}^p$ is the gust disturbance input, and $\mathbf{F}_{NR}(\mathbf{x})$ contains the nonlinear terms from the Taylor series expansion. The matrices $\mathbf{A}$, $\mathbf{B}_c$, and $\mathbf{B}_g$ are assumed known from the model reduction procedure, while the nonlinear function $\mathbf{F}_{NR}$ is available for simulation but is not directly used in the controller design (which is based on the linear part of the model).

\subsection{Reference Model Definition and Selection}
\label{sec:mrac_refmodel}

The reference model specifies the desired closed-loop behaviour that the adaptive controller should achieve. It takes the form:
\begin{equation}
    \dot{\mathbf{x}}_m(t) = \mathbf{A}_m\,\mathbf{x}_m(t) + \mathbf{B}_m\,\mathbf{r}(t) + \mathbf{B}_{g}\,\mathbf{u}_d(t) + \mathbf{F}_{NR}(\mathbf{x}_m(t))
    \label{eq:refmodel}
\end{equation}
where $\mathbf{x}_m(t) \in \mathbb{R}^n$ is the reference model state, $\mathbf{r}(t) \in \mathbb{R}^m$ is the reference command, and $\mathbf{A}_m \in \mathbb{R}^{n \times n}$ is a Hurwitz matrix that defines the desired dynamics. The reference model shares the same gust input matrix $\mathbf{B}_g$ and nonlinear function $\mathbf{F}_{NR}$ as the plant, so that the adaptive controller need only compensate for the difference in linear dynamics.

The design of $\mathbf{A}_m$ is the key step in reference model selection. It is constructed by modifying the damping ratios of the dominant aeroelastic modes in the open-loop system matrix $\mathbf{A}$. Specifically, each complex-conjugate eigenvalue pair $\lambda_k = -\sigma_k \pm j\omega_{d,k}$ of $\mathbf{A}$ is replaced by:
\begin{equation}
    \lambda_{m,k} = -\sigma_{m,k} \pm j\omega_{d,m,k}
    \label{eq:refmodel_eigs}
\end{equation}
where $\sigma_{m,k} > \sigma_k$, corresponding to an increased damping ratio $\zeta_{m,k} > \zeta_k$. The natural frequency can be kept the same or modified as needed. The reference model matrix is then:
\begin{equation}
    \mathbf{A}_m = \mathrm{diag}(\lambda_{m,1}, \lambda_{m,1}^*, \ldots, \lambda_{m,n/2}, \lambda_{m,n/2}^*)
    \label{eq:Am_construction}
\end{equation}

The choice of reference model damping involves a trade-off. Higher damping ratios in the reference model lead to faster decay of the gust response and larger load reductions, but also require larger control surface deflections and higher adaptive gain magnitudes. In practice, the reference model damping is increased by a factor of 2--5 relative to the open-loop values for the dominant aeroelastic modes, while keeping the flight-dynamic modes (phugoid, short period) at their open-loop values to avoid interfering with the aircraft's handling qualities~\citep{Tantaroudas2017bookchapter}.

\subsection{Adaptive Control Law}
\label{sec:mrac_control}

The control law takes the form:
\begin{equation}
    \mathbf{u}_c(t) = \boldsymbol{\theta}^T(t)\,\boldsymbol{\phi}(t) = \mathbf{K}_x(t)\,\mathbf{x}(t) + \mathbf{K}_r(t)\,\mathbf{r}(t)
    \label{eq:control_law}
\end{equation}
where $\boldsymbol{\theta}(t) = [\mathbf{K}_x(t)^T, \mathbf{K}_r(t)^T]^T \in \mathbb{R}^{(n+m) \times m}$ is the matrix of time-varying adaptive gains and $\boldsymbol{\phi}(t) = [\mathbf{x}(t)^T, \mathbf{r}(t)^T]^T \in \mathbb{R}^{n+m}$ is the regression vector. The gain matrices $\mathbf{K}_x(t) \in \mathbb{R}^{n \times m}$ and $\mathbf{K}_r(t) \in \mathbb{R}^{m \times m}$ are updated at each time step by the adaptation law (derived below).

The ideal (unknown) gains $\mathbf{K}_x^*$ and $\mathbf{K}_r^*$ are defined as those that achieve perfect model matching in the linear case:
\begin{equation}
    \mathbf{A} + \mathbf{B}_c\,\mathbf{K}_x^* = \mathbf{A}_m
    \label{eq:matching_Kx}
\end{equation}
\begin{equation}
    \mathbf{B}_c\,\mathbf{K}_r^* = \mathbf{B}_m
    \label{eq:matching_Kr}
\end{equation}

These model matching conditions require that the pair $(\mathbf{A}, \mathbf{B}_c)$ is controllable and that $\mathbf{B}_c$ has sufficient rank. The existence of $\mathbf{K}_x^*$ and $\mathbf{K}_r^*$ is a necessary condition for MRAC; when these conditions are not satisfied exactly, approximate model matching can still be employed with bounded residual error.

\subsection{Tracking Error Dynamics}
\label{sec:mrac_error}

The tracking error is defined as the difference between the plant and reference model states:
\begin{equation}
    \mathbf{e}(t) = \mathbf{x}(t) - \mathbf{x}_m(t)
    \label{eq:error_def}
\end{equation}

The parameter error is the difference between the current and ideal gains:
\begin{equation}
    \tilde{\boldsymbol{\theta}}(t) = \boldsymbol{\theta}(t) - \boldsymbol{\theta}^*
    \label{eq:param_error}
\end{equation}
where $\boldsymbol{\theta}^* = [\mathbf{K}_x^{*T}, \mathbf{K}_r^{*T}]^T$ is the ideal gain matrix.

Subtracting the reference model equation~\eqref{eq:refmodel} from the plant equation~\eqref{eq:plant}, and using the model matching conditions~\eqref{eq:matching_Kx}--\eqref{eq:matching_Kr} and the control law~\eqref{eq:control_law}, the error dynamics are obtained:
\begin{equation}
    \dot{\mathbf{e}}(t) = \mathbf{A}_m\,\mathbf{e}(t) + \mathbf{B}_c\,\tilde{\boldsymbol{\theta}}^T(t)\,\boldsymbol{\phi}(t) + \mathbf{F}_{Df}(\mathbf{x}, \mathbf{x}_m)
    \label{eq:error_dynamics}
\end{equation}
where $\mathbf{F}_{Df}(\mathbf{x}, \mathbf{x}_m) = \mathbf{F}_{NR}(\mathbf{x}) - \mathbf{F}_{NR}(\mathbf{x}_m)$ is the differential nonlinearity, the difference in the nonlinear function evaluated at the plant and reference model states.

In the purely linear case ($\mathbf{F}_{NR} = \mathbf{0}$), the error dynamics reduce to:
\begin{equation}
    \dot{\mathbf{e}}(t) = \mathbf{A}_m\,\mathbf{e}(t) + \mathbf{B}_c\,\tilde{\boldsymbol{\theta}}^T(t)\,\boldsymbol{\phi}(t)
    \label{eq:error_linear}
\end{equation}
which is the standard form for MRAC error dynamics. The objective of the adaptation law is to drive $\tilde{\boldsymbol{\theta}}(t) \to \mathbf{0}$ (or at least to ensure $\mathbf{e}(t) \to \mathbf{0}$) as $t \to \infty$.

\subsection{Lyapunov Stability Analysis}
\label{sec:mrac_lyapunov}

The stability of the MRAC system is established using a Lyapunov approach. This analysis serves two purposes: it provides a constructive proof of stability, and it yields the adaptation law as a by-product of the stability requirement.

\subsubsection{Lyapunov Function Candidate}

Consider the positive-definite function:
\begin{equation}
    V(\mathbf{e}, \tilde{\boldsymbol{\theta}}) = \mathbf{e}^T\mathbf{P}\,\mathbf{e} + \mathrm{tr}\!\left(\tilde{\boldsymbol{\theta}}^T\boldsymbol{\Gamma}^{-1}\tilde{\boldsymbol{\theta}}\right)
    \label{eq:lyapunov}
\end{equation}
where $\mathbf{P} \in \mathbb{R}^{n \times n}$ is a symmetric positive-definite matrix and $\boldsymbol{\Gamma} \in \mathbb{R}^{(n+m) \times (n+m)}$ is a symmetric positive-definite matrix known as the adaptation rate (or learning rate) matrix. The first term $\mathbf{e}^T\mathbf{P}\mathbf{e}$ penalises the tracking error, while the second term $\mathrm{tr}(\tilde{\boldsymbol{\theta}}^T\boldsymbol{\Gamma}^{-1}\tilde{\boldsymbol{\theta}})$ penalises the parameter estimation error. The function $V$ is positive definite in $(\mathbf{e}, \tilde{\boldsymbol{\theta}})$ and radially unbounded, satisfying the requirements for a Lyapunov function candidate.

The matrix $\mathbf{P}$ is obtained as the unique positive-definite solution of the Lyapunov equation:
\begin{equation}
    \mathbf{A}_m^T\mathbf{P} + \mathbf{P}\,\mathbf{A}_m = -\mathbf{Q}
    \label{eq:lyap_eq}
\end{equation}
where $\mathbf{Q} \in \mathbb{R}^{n \times n}$ is a user-selected symmetric positive-definite (or positive semi-definite) matrix. Since $\mathbf{A}_m$ is Hurwitz by construction, a unique positive-definite $\mathbf{P}$ exists for any positive-definite $\mathbf{Q}$~\citep{Tantaroudas2017bookchapter}.

\subsubsection{Time Derivative of the Lyapunov Function}

Differentiating $V$ along the trajectories of the error system~\eqref{eq:error_dynamics} (and noting that $\dot{\tilde{\boldsymbol{\theta}}} = \dot{\boldsymbol{\theta}}$ since $\boldsymbol{\theta}^*$ is constant):
\begin{align}
    \dot{V} &= \dot{\mathbf{e}}^T\mathbf{P}\,\mathbf{e} + \mathbf{e}^T\mathbf{P}\,\dot{\mathbf{e}} + 2\,\mathrm{tr}\!\left(\tilde{\boldsymbol{\theta}}^T\boldsymbol{\Gamma}^{-1}\dot{\boldsymbol{\theta}}\right) \nonumber \\
    &= \mathbf{e}^T(\mathbf{A}_m^T\mathbf{P} + \mathbf{P}\,\mathbf{A}_m)\mathbf{e} + 2\,\mathbf{e}^T\mathbf{P}\,\mathbf{B}_c\,\tilde{\boldsymbol{\theta}}^T\boldsymbol{\phi} + 2\,\mathrm{tr}\!\left(\tilde{\boldsymbol{\theta}}^T\boldsymbol{\Gamma}^{-1}\dot{\boldsymbol{\theta}}\right) + 2\,\mathbf{e}^T\mathbf{P}\,\mathbf{F}_{Df}
    \label{eq:Vdot_expanded}
\end{align}

Substituting the Lyapunov equation~\eqref{eq:lyap_eq} and using the trace identity $\mathbf{e}^T\mathbf{P}\mathbf{B}_c\tilde{\boldsymbol{\theta}}^T\boldsymbol{\phi} = \mathrm{tr}(\tilde{\boldsymbol{\theta}}^T\boldsymbol{\phi}\,\mathbf{e}^T\mathbf{P}\,\mathbf{B}_c)$:
\begin{equation}
    \dot{V} = -\mathbf{e}^T\mathbf{Q}\,\mathbf{e} + 2\,\mathrm{tr}\!\left(\tilde{\boldsymbol{\theta}}^T\!\left[\boldsymbol{\Gamma}^{-1}\dot{\boldsymbol{\theta}} + \boldsymbol{\phi}\,\mathbf{e}^T\mathbf{P}\,\mathbf{B}_c\right]\right) + 2\,\mathbf{e}^T\mathbf{P}\,\mathbf{F}_{Df}
    \label{eq:Vdot_simplified}
\end{equation}

\subsubsection{Derivation of the Adaptation Law}

The second term in~\eqref{eq:Vdot_simplified} involves the unknown parameter error $\tilde{\boldsymbol{\theta}}$. To eliminate this term and ensure $\dot{V} \leq 0$ (at least in the linear case), we choose the adaptation law such that the expression in square brackets vanishes:
\begin{equation}
    \boldsymbol{\Gamma}^{-1}\dot{\boldsymbol{\theta}} + \boldsymbol{\phi}\,\mathbf{e}^T\mathbf{P}\,\mathbf{B}_c = \mathbf{0}
\end{equation}
which yields the adaptation law:
\begin{equation}
    \boxed{\dot{\boldsymbol{\theta}}(t) = -\boldsymbol{\Gamma}\,\boldsymbol{\phi}(t)\,\mathbf{e}^T(t)\,\mathbf{P}\,\mathbf{B}_c}
    \label{eq:adapt_law}
\end{equation}

With this choice, the time derivative of the Lyapunov function becomes:
\begin{equation}
    \dot{V} = -\mathbf{e}^T\mathbf{Q}\,\mathbf{e} + 2\,\mathbf{e}^T\mathbf{P}\,\mathbf{F}_{Df}(\mathbf{x}, \mathbf{x}_m)
    \label{eq:Vdot_final}
\end{equation}

\subsubsection{Linear Case: Asymptotic Stability}

In the linear case ($\mathbf{F}_{NR} = \mathbf{0}$, hence $\mathbf{F}_{Df} = \mathbf{0}$), the Lyapunov derivative reduces to:
\begin{equation}
    \dot{V} = -\mathbf{e}^T\mathbf{Q}\,\mathbf{e} \leq 0
    \label{eq:Vdot_linear}
\end{equation}

Since $V > 0$ and $\dot{V} \leq 0$, the origin $(\mathbf{e}, \tilde{\boldsymbol{\theta}}) = (\mathbf{0}, \mathbf{0})$ is stable in the sense of Lyapunov. Moreover, $\dot{V} = 0$ only when $\mathbf{e} = \mathbf{0}$, and by Barbalat's lemma (noting that $\ddot{V}$ is bounded for bounded signals), $\mathbf{e}(t) \to \mathbf{0}$ as $t \to \infty$. This establishes asymptotic tracking: the plant state converges to the reference model state regardless of the initial parameter error, provided the model matching conditions are satisfied.

Note that $\dot{V} \leq 0$ does not imply $\tilde{\boldsymbol{\theta}}(t) \to \mathbf{0}$; the adaptive gains converge to values that achieve zero tracking error, but these need not equal the ideal gains $\boldsymbol{\theta}^*$. This is a well-known property of MRAC: parameter convergence requires persistent excitation, which is generally not guaranteed in a GLA application where the gust disturbance is transient.

\subsection{Stability Condition for Nonlinear Terms}
\label{sec:mrac_nonlinear}

In the nonlinear case, the differential nonlinearity $\mathbf{F}_{Df}$ contributes to $\dot{V}$ through the term $2\,\mathbf{e}^T\mathbf{P}\,\mathbf{F}_{Df}$. Applying the Cauchy--Schwarz inequality:
\begin{equation}
    2\,\mathbf{e}^T\mathbf{P}\,\mathbf{F}_{Df} \leq 2\,\|\mathbf{P}\|\,\|\mathbf{e}\|\,\|\mathbf{F}_{Df}\|
    \label{eq:cauchy_schwarz}
\end{equation}
where $\|\cdot\|$ denotes the spectral (induced 2-)norm. For $\dot{V} \leq 0$, we require:
\begin{equation}
    -\mathbf{e}^T\mathbf{Q}\,\mathbf{e} + 2\,\|\mathbf{P}\|\,\|\mathbf{e}\|\,\|\mathbf{F}_{Df}\| \leq 0
\end{equation}

Since $\mathbf{e}^T\mathbf{Q}\,\mathbf{e} \geq \lambda_{\min}(\mathbf{Q})\,\|\mathbf{e}\|^2$, a sufficient condition is:
\begin{equation}
    \|\mathbf{F}_{NR}(\mathbf{x}) - \mathbf{F}_{NR}(\mathbf{x}_m)\| \leq \frac{\lambda_{\min}(\mathbf{Q})}{2\,\|\mathbf{P}\|}\,\|\mathbf{x} - \mathbf{x}_m\|
    \label{eq:lipschitz}
\end{equation}

This is a Lipschitz condition on the nonlinear function $\mathbf{F}_{NR}$: it requires that the nonlinear terms do not grow faster than the stabilising effect of the reference model. The Lipschitz constant $L_F = \lambda_{\min}(\mathbf{Q})/(2\|\mathbf{P}\|)$ depends on the reference model design (through $\mathbf{A}_m$, which determines $\mathbf{P}$ and $\mathbf{Q}$) and can be increased by choosing a reference model with higher damping.

For typical aeroelastic systems with polynomial (cubic hardening) nonlinearities, the Lipschitz condition~\eqref{eq:lipschitz} is satisfied in a neighbourhood of the equilibrium whose size depends on the nonlinearity strength and the reference model damping~\citep{Tantaroudas2014aviation, Tantaroudas2017bookchapter}. Outside this neighbourhood, the adaptive controller may still provide load alleviation, but the formal stability guarantee no longer holds.

\subsection{Role of the Adaptation Rate Matrix \texorpdfstring{$\boldsymbol{\Gamma}$}{Gamma}}
\label{sec:mrac_gamma}

The adaptation rate matrix $\boldsymbol{\Gamma}$ is the single most important design parameter in the MRAC formulation. Its effect can be understood from the adaptation law~\eqref{eq:adapt_law}: $\boldsymbol{\Gamma}$ scales the rate at which the adaptive gains $\boldsymbol{\theta}$ are updated in response to the tracking error. Regarding convergence speed, larger $\boldsymbol{\Gamma}$ leads to faster gain adaptation and faster convergence of the tracking error $\mathbf{e}(t) \to \mathbf{0}$; in the Lyapunov function~\eqref{eq:lyapunov}, a larger $\boldsymbol{\Gamma}$ reduces the ``cost'' of parameter error relative to tracking error, allowing the system to prioritise tracking performance. In terms of sensitivity to disturbances, a larger $\boldsymbol{\Gamma}$ causes the gains to respond more aggressively to the gust-induced tracking error, leading to larger control surface deflections and greater load alleviation, although excessively large $\boldsymbol{\Gamma}$ can cause gain oscillations and actuator saturation. Conversely, smaller $\boldsymbol{\Gamma}$ provides more robust behaviour in the presence of unmodelled dynamics or sensor noise, since the gains change slowly and are less susceptible to high-frequency perturbations. In this work, $\boldsymbol{\Gamma}$ is parameterised as a scalar multiple of the weighting matrix: $\boldsymbol{\Gamma} = \gamma\,\mathbf{Q}$, where $\gamma > 0$ is a scalar, reducing the design to a single parameter and ensuring dimensional consistency. Values of $\gamma$ ranging from 0.01 to 1.0 are investigated.

\subsection{Treatment of Non-Minimum Phase Dynamics}
\label{sec:mrac_nmp}

A fundamental requirement for direct MRAC is that the plant transfer function from the control input to the output must be minimum phase, that is, all transmission zeros must lie in the open left half-plane. When the plant has unstable (right-half-plane) zeros, the direct adaptive law~\eqref{eq:adapt_law} can lead to unbounded signals.

For the aeroelastic systems considered in this work, the state-space representation obtained from the NMOR may exhibit non-minimum phase behaviour due to the coupling between structural and aerodynamic states. This is addressed using the Bass--Gura formula~\citep{Tantaroudas2017bookchapter, DaRonch2013control}: a preliminary state-feedback controller is designed to place the unstable zeros in the left half-plane, producing a modified plant that is minimum phase. The adaptive law is then applied to this modified plant. The preliminary controller is designed using only the linear part of the ROM and does not affect the stability proof, since it amounts to a change of coordinates in the state space.

The procedure is as follows. First, the transmission zeros of the linear ROM transfer function $\mathbf{C}(s\mathbf{I} - \mathbf{A})^{-1}\mathbf{B}_c$ are computed. If any zeros have positive real parts, a state-feedback gain $\mathbf{K}_0$ is designed using the Bass--Gura formula to relocate them to the left half-plane. The modified plant is then defined as $\dot{\mathbf{x}} = (\mathbf{A} + \mathbf{B}_c\mathbf{K}_0)\mathbf{x} + \mathbf{B}_c\mathbf{v} + \mathbf{B}_g\mathbf{u}_d + \mathbf{F}_{NR}(\mathbf{x})$, where $\mathbf{v}$ is the new control input. The MRAC design is then applied to this modified plant with the adaptive control law $\mathbf{v}(t) = \mathbf{K}_x(t)\mathbf{x}(t) + \mathbf{K}_r(t)\mathbf{r}(t)$, and the total control input is $\mathbf{u}_c = \mathbf{K}_0\mathbf{x} + \mathbf{v}$.

\subsection{Implementation Considerations}
\label{sec:mrac_implementation}

The MRAC controller is implemented in discrete time using a fourth-order Runge--Kutta integration scheme with time step $\Delta t$ matched to the aeroelastic simulation. At each time step, the plant state $\mathbf{x}(t)$ is measured (or estimated), the reference model~\eqref{eq:refmodel} is integrated to obtain $\mathbf{x}_m(t)$, and the tracking error $\mathbf{e}(t) = \mathbf{x}(t) - \mathbf{x}_m(t)$ is computed. The adaptive gains are then updated using the discretised adaptation law~\eqref{eq:adapt_law}, the control input $\mathbf{u}_c(t)$ is computed from~\eqref{eq:control_law}, and $\mathbf{u}_c(t)$ is applied to the plant to advance the simulation.

The computational cost of the MRAC controller is negligible compared to the aeroelastic simulation, since all operations involve matrices of dimension $n$ (the ROM size, typically 8--16 states). This makes the approach suitable for real-time implementation.

%% ============================================================
\section{Results: Three-Degree-of-Freedom Aerofoil}
\label{sec:3dof}
%% ============================================================

\subsection{Test Case Description}
\label{sec:3dof_description}

The first test case is a pitch--plunge--flap aerofoil, a classical benchmark for nonlinear aeroelastic analysis~\citep{Alighanbari1995, Irani2011, Dowell2004}. The system has three structural degrees of freedom, pitch ($\alpha$), plunge ($h$), and trailing-edge flap deflection ($\beta$), coupled through the unsteady aerodynamic forces. The aerodynamics are modelled using Theodorsen's theory~\citep{Theodorsen1935} with Jones' rational approximation~\citep{Jones1938} for the Wagner function~\citep{Wagner1925}, augmented by K\"{u}ssner's function~\citep{Kussner1936} for gust response.

Cubic stiffness nonlinearities are included in the pitch and plunge degrees of freedom:
\begin{equation}
    K_\alpha(\alpha) = K_{\alpha,1}\,\alpha + K_{\alpha,3}\,\alpha^3, \qquad
    K_\xi(\xi) = K_{\xi,1}\,\xi + K_{\xi,3}\,\xi^3
    \label{eq:cubic_stiffness}
\end{equation}
with hardening parameters $K_{\alpha,3} = 3.0$ and $K_{\xi,3} = 1.0$. These values produce limit cycle oscillations (LCOs) above the linear flutter speed, providing a challenging test for the adaptive controller.

The full aeroelastic model has $N = 14$ states (3 structural positions, 3 structural velocities, and 8 aerodynamic lag states). The NMOR retains $n = 8$ reduced states: three complex-conjugate eigenvalue pairs corresponding to the three structural degrees of freedom (pitch, plunge, and flap) and two real eigenvalues equal to the K\"{u}ssner constant $\epsilon_3 = -0.1393$, which capture the gust influence on the aerodynamic response~\citep{DaRonch2013gust}.

\subsection{Nonlinear ROM Validation}
\label{sec:3dof_nrom}

\begin{figure}[htbp]
    \centering
    \begin{subfigure}[b]{0.48\textwidth}
        \centering
        \includegraphics[width=\textwidth]{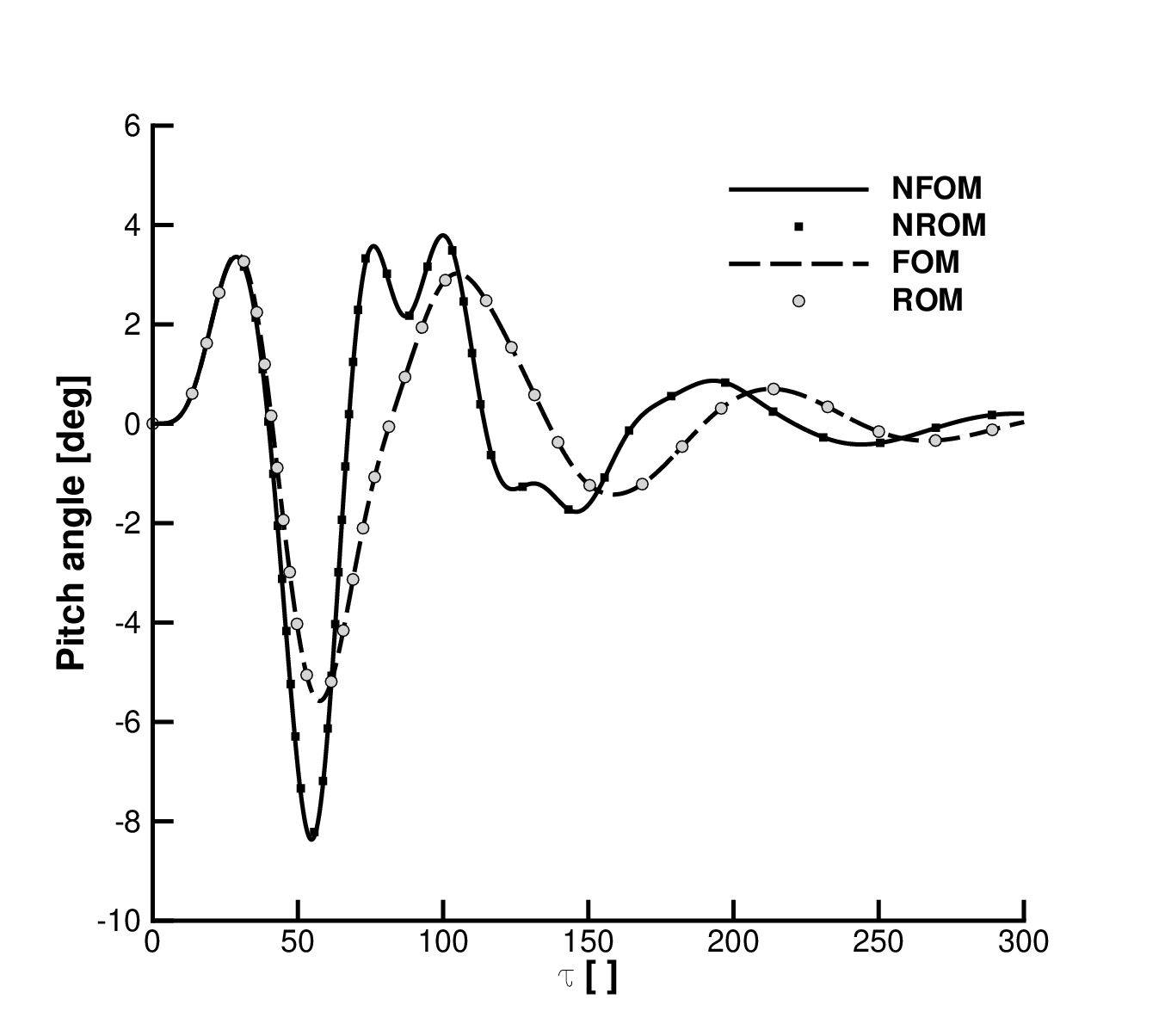}
        \caption{Pitch angle $\alpha(t)$.}
        \label{fig:nrom_pitch}
    \end{subfigure}
    \hfill
    \begin{subfigure}[b]{0.48\textwidth}
        \centering
        \includegraphics[width=\textwidth]{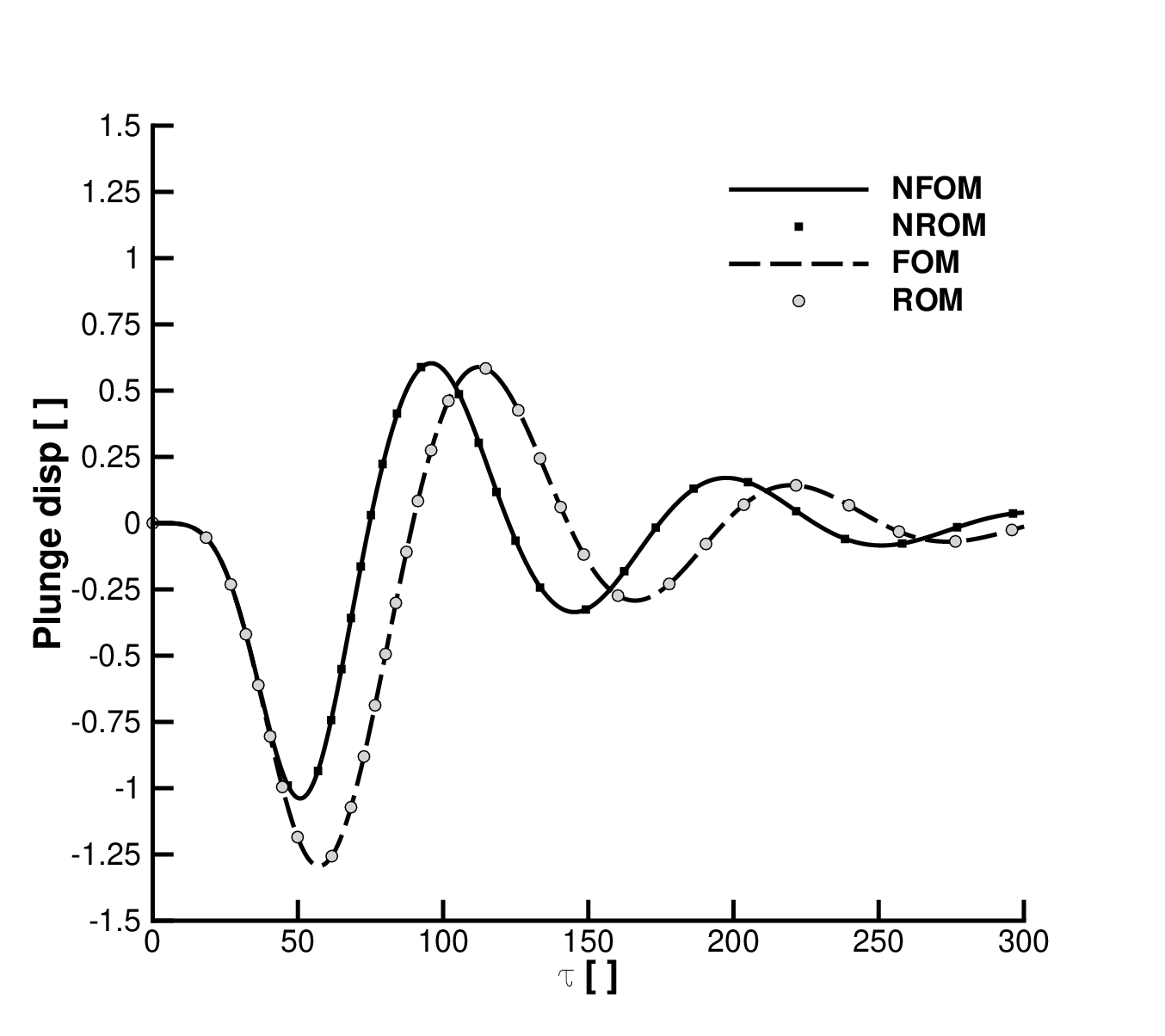}
        \caption{Plunge displacement $h(t)$.}
        \label{fig:nrom_plunge}
    \end{subfigure}

    \vspace{0.5cm}
    \begin{subfigure}[b]{0.48\textwidth}
        \centering
        \includegraphics[width=\textwidth]{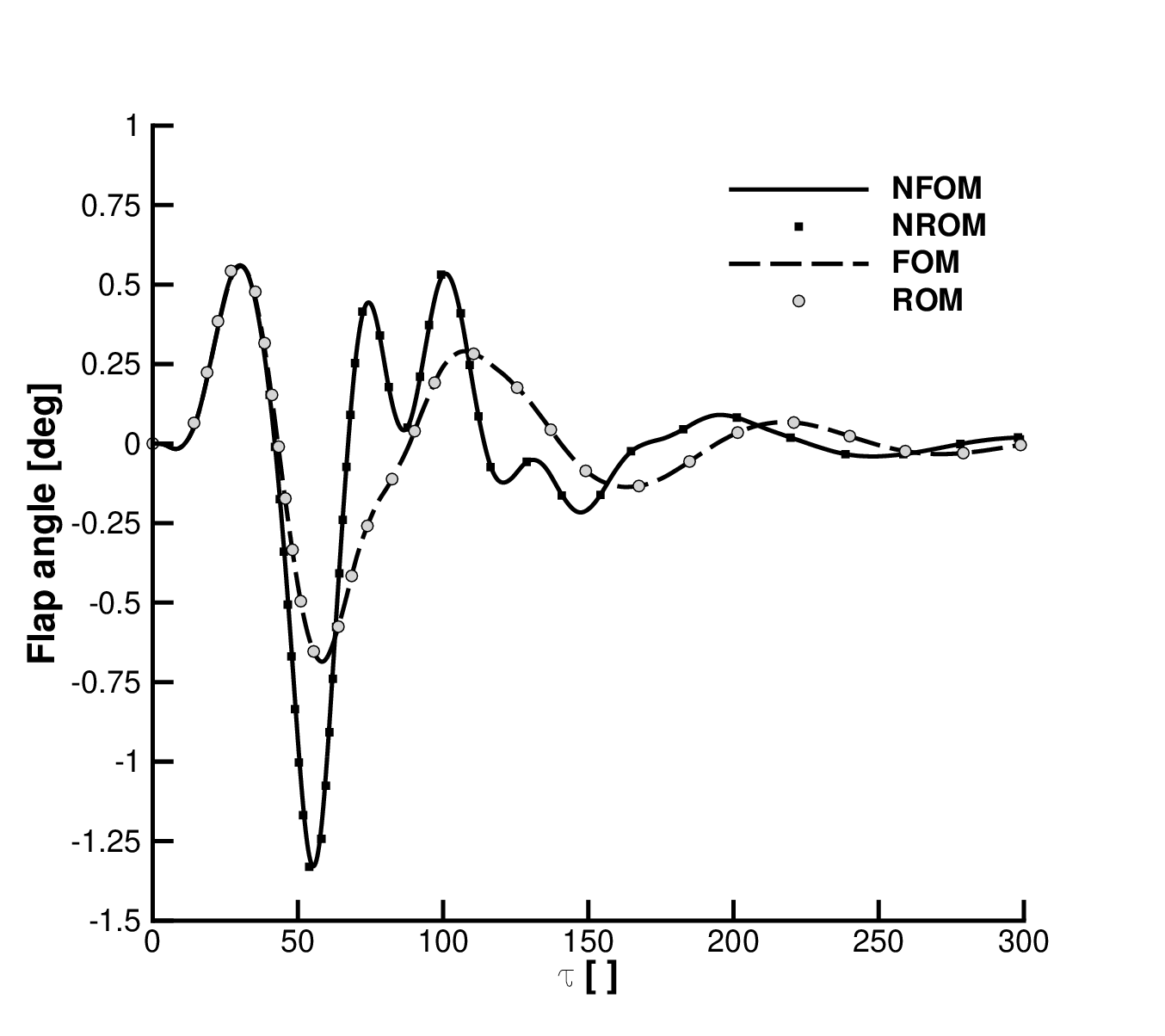}
        \caption{Flap deflection $\delta(t)$.}
        \label{fig:nrom_delta}
    \end{subfigure}
    \caption{Aeroelastic response at $U^* = 4.5$ for the worst-case ``1-cosine'' gust of intensity $W_0 = 0.14$: comparison of the nonlinear full-order model (NFOM) against the linear full-order model and the nonlinear reduced-order model (NROM) for the 3-DOF aerofoil.}
    \label{fig:nrom_validation}
\end{figure}

The accuracy of the eight-state NROM is assessed by comparing the aeroelastic response under the worst-case ``1-cosine'' gust (with intensity $W_0 = 0.14$ of the freestream speed at $U^* = 4.5$) against the nonlinear full-order model and the corresponding linear models. \Cref{fig:nrom_validation} shows the pitch, plunge, and flap deflection time histories for this worst-case gust of gradient length $H_g = 55$ semichords.

The eight-state NROM provides excellent agreement with the nonlinear full-order model for both frequency and amplitude of the gust response. The nonlinear ROM accurately captures the differences between the linear and nonlinear responses, which become more pronounced at larger gust amplitudes due to the cubic stiffness nonlinearities. This level of accuracy is sufficient for control design and validation.

\subsection{MRAC Design for the 3-DOF Aerofoil}
\label{sec:3dof_mrac_design}

The reference model is designed by increasing the damping ratios of the three retained complex-conjugate mode pairs relative to their open-loop values, while the two real gust eigenvalues are kept at their open-loop values (equal to the K\"{u}ssner constant). The matrix $\mathbf{Q}$ is chosen as a positive-definite diagonal matrix with elements $Q_{11} = Q_{22} = Q_{66} = 10$ and $Q_{33} = Q_{44} = Q_{55} = Q_{77} = Q_{88} = 30$, following~\citet{Tantaroudas2017bookchapter}. Three adaptation rates are tested: $\boldsymbol{\Gamma} = 0.1\,\mathbf{Q}$, $0.5\,\mathbf{Q}$, and $1.0\,\mathbf{Q}$.

The gust input is a ``1-cosine'' discrete gust with gradient length $H_g = 55$ semichords and 14\% freestream velocity intensity, representing a worst-case certification gust~\citep{DaRonch2013gust, Fichera2014isma}.

\subsection{Gust Load Alleviation Results}
\label{sec:3dof_results}

\begin{figure}[htbp]
    \centering
    \begin{subfigure}[b]{0.48\textwidth}
        \centering
        \includegraphics[width=\textwidth]{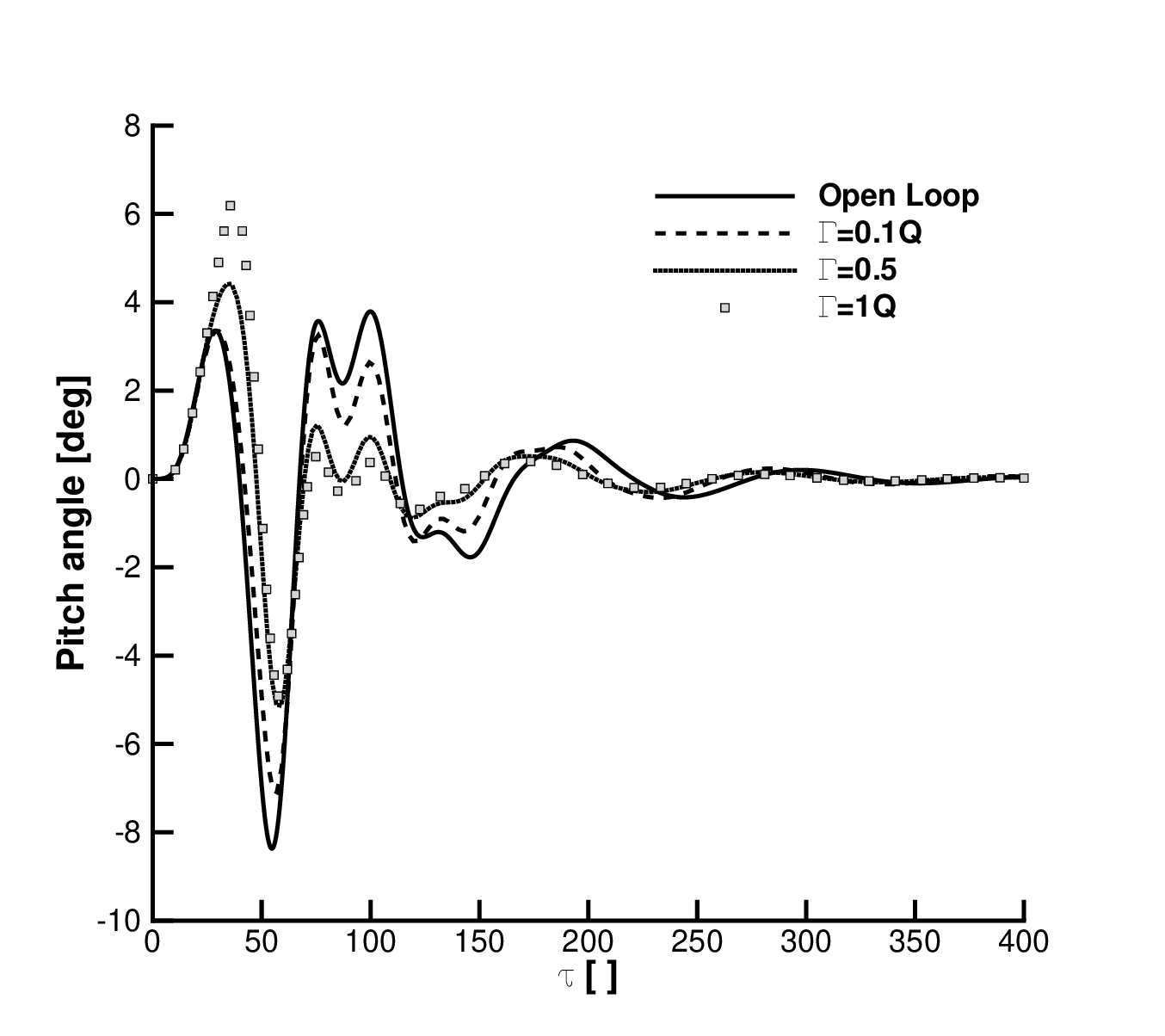}
        \caption{Pitch angle $\alpha(t)$.}
        \label{fig:pitch_gla}
    \end{subfigure}
    \hfill
    \begin{subfigure}[b]{0.48\textwidth}
        \centering
        \includegraphics[width=\textwidth]{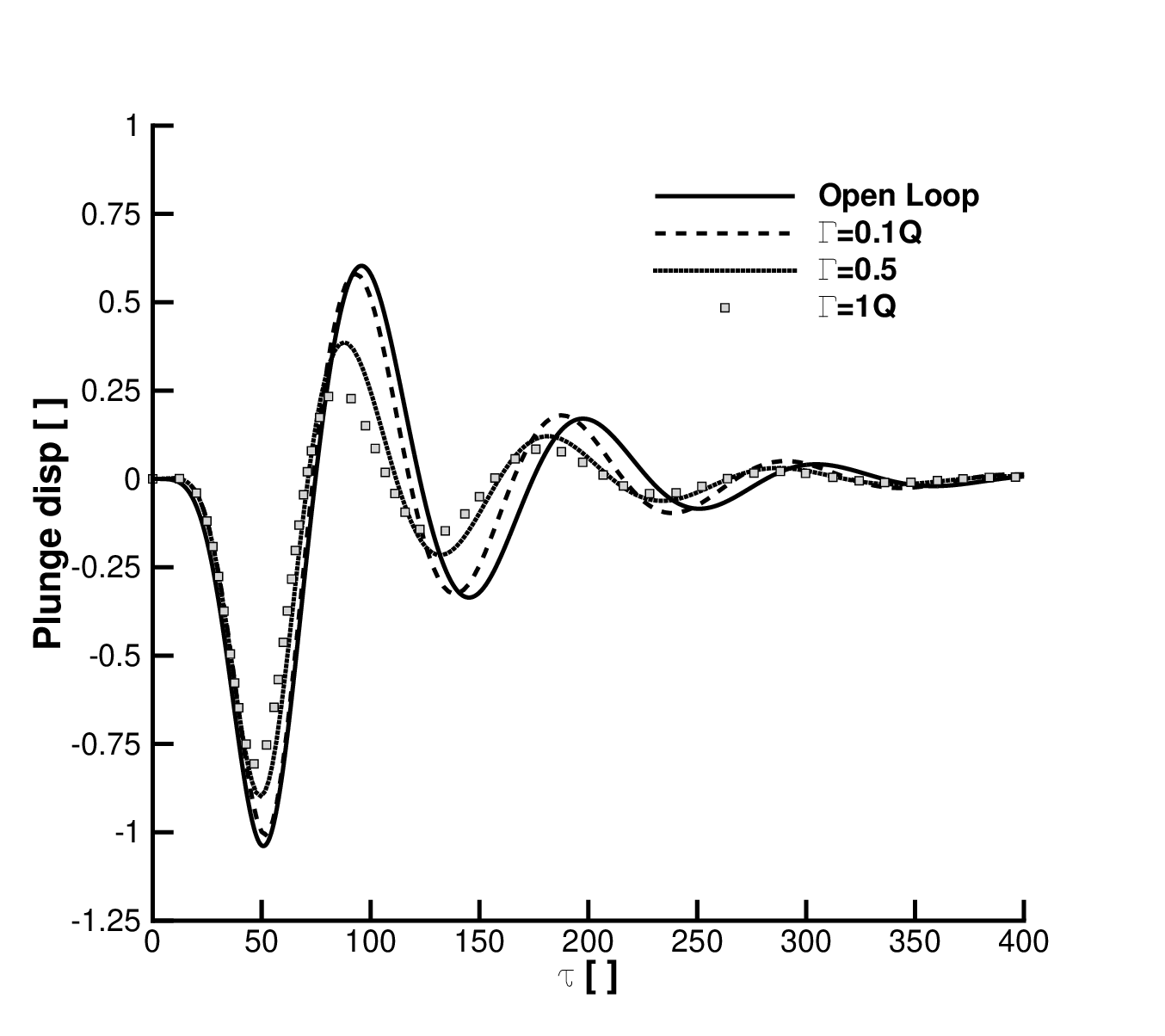}
        \caption{Plunge displacement $h(t)$.}
        \label{fig:plunge_gla}
    \end{subfigure}

    \vspace{0.5cm}
    \begin{subfigure}[b]{0.48\textwidth}
        \centering
        \includegraphics[width=\textwidth]{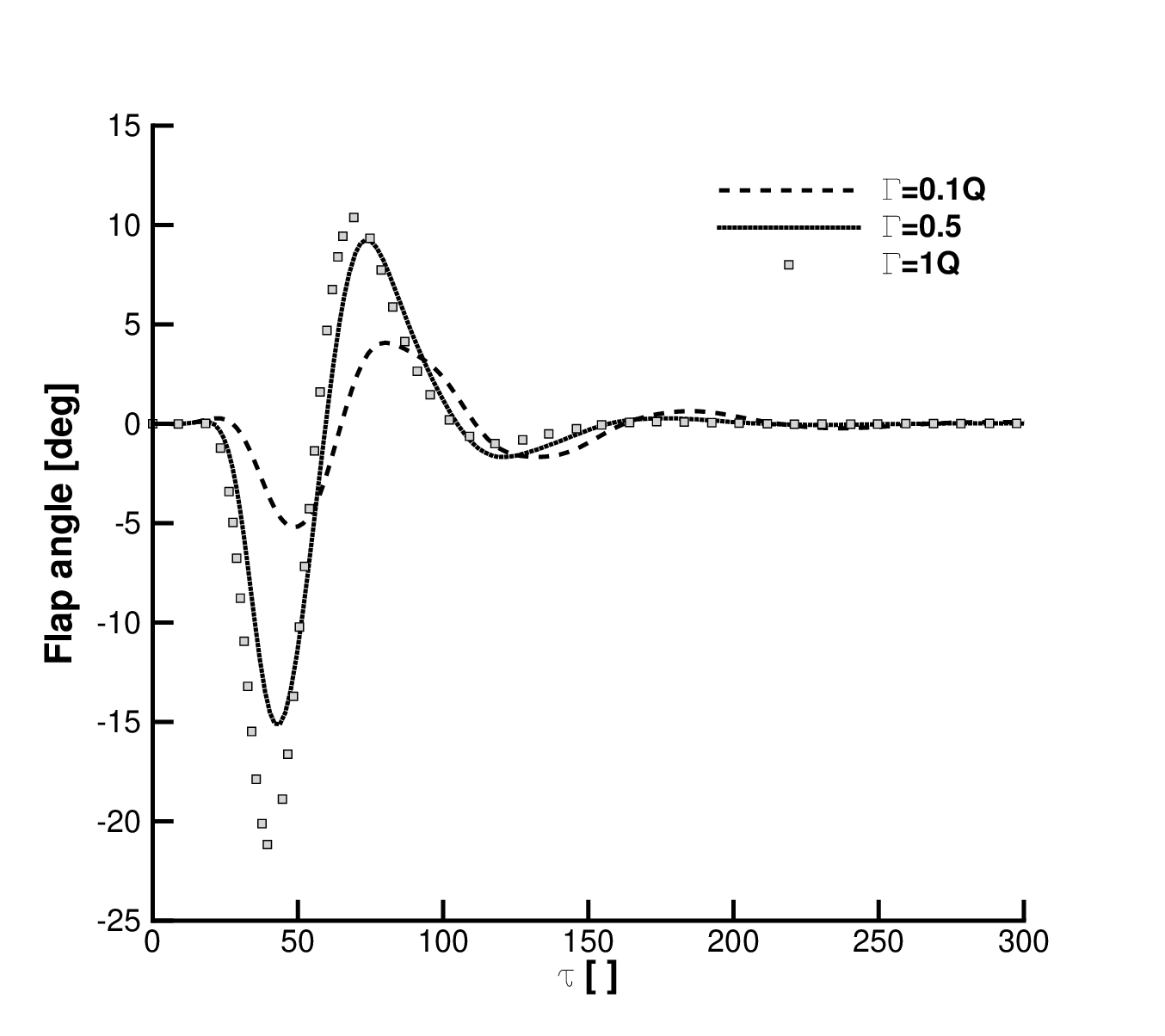}
        \caption{Flap deflection $\delta(t)$ (control input).}
        \label{fig:flap_gla}
    \end{subfigure}
    \hfill
    \begin{subfigure}[b]{0.48\textwidth}
        \centering
        \includegraphics[width=\textwidth]{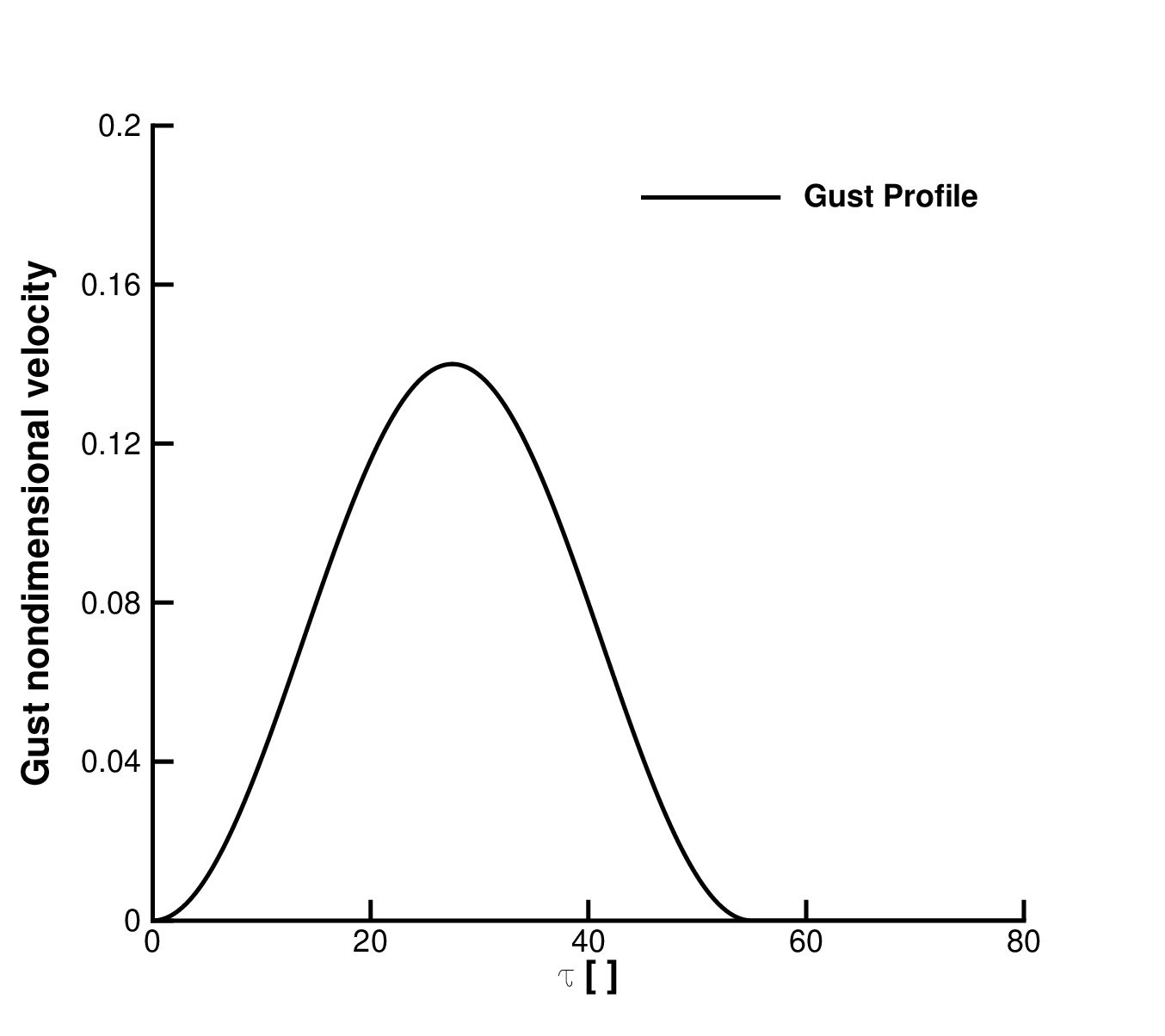}
        \caption{Gust velocity profile $w_g(t)$.}
        \label{fig:gust_gla}
    \end{subfigure}
    \caption{Gust load alleviation for the 3-DOF aerofoil under worst-case discrete gust: open-loop response and MRAC-controlled response at three adaptation rates ($\boldsymbol{\Gamma} = 0.1\,\mathbf{Q}$, $0.5\,\mathbf{Q}$, $1.0\,\mathbf{Q}$).}
    \label{fig:3dof_gla}
\end{figure}

\Cref{fig:3dof_gla} presents the pitch, plunge, and flap deflection time histories under the worst-case gust for the open-loop system and the MRAC-controlled system at three adaptation rates.

The key observations from \cref{fig:3dof_gla} are as follows. At the low adaptation rate ($\boldsymbol{\Gamma} = 0.1\,\mathbf{Q}$), the gains adapt slowly, resulting in a gradual reduction of the oscillation amplitude, with flap deflections that are smooth and well within actuator limits. At the moderate adaptation rate ($\boldsymbol{\Gamma} = 0.5\,\mathbf{Q}$), the gains adapt more quickly, achieving a more significant reduction in both pitch and plunge responses; the flap deflections show slightly larger initial transients but remain within practical bounds. At the high adaptation rate ($\boldsymbol{\Gamma} = 1.0\,\mathbf{Q}$), the fastest adaptation achieves the largest load reduction, but the initial pitch response shows a brief overshoot before the gains converge, and the flap deflections are the largest of the three cases, consistent with the increased adaptation effort. These results confirm the trade-off between load reduction performance and actuator demand that is governed by the adaptation rate $\boldsymbol{\Gamma}$. All three controllers successfully alleviate the gust loads while maintaining bounded signals and stable adaptation.

%% ============================================================
\section{Results: Global Hawk-Like UAV}
\label{sec:uav}
%% ============================================================

\subsection{Vehicle Description}
\label{sec:uav_description}

\begin{figure}[htbp]
    \centering
    \includegraphics[width=0.65\textwidth]{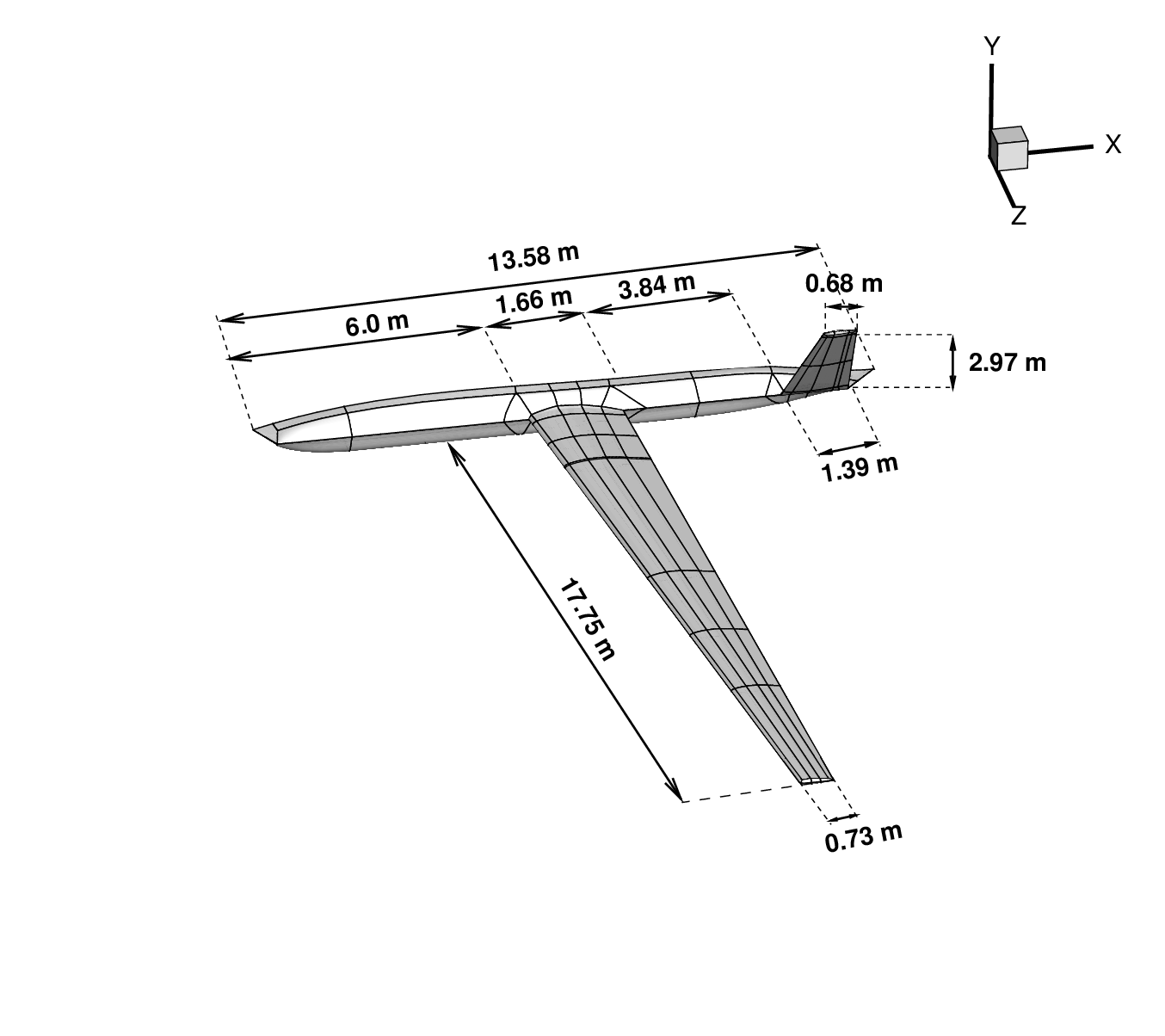}
    \caption{Geometry of the Global Hawk-like UAV model showing the high-aspect-ratio wing, fuselage, and tail surfaces. Control surfaces (trailing-edge flaps) are distributed between 37\% and 77\% of the wing span on each semi-wing.}
    \label{fig:dstl_geometry}
\end{figure}

The second test case is a Global Hawk-like UAV model developed at the University of Southampton~\citep{Tantaroudas2014aviation, Tantaroudas2017bookchapter, DaRonch2014scitech_flight}. The aircraft (\cref{fig:dstl_geometry}) has a high-aspect-ratio wing with significant structural flexibility, representative of HALE platforms. The structural model uses a geometrically exact beam formulation~\citep{Hodges2003, Palacios2010} with $N = 540$ degrees of freedom. The unsteady aerodynamics are computed using a thin-strip theory formulation based on two-dimensional thin-aerofoil theory~\citep{Theodorsen1935}. The coupled aeroelastic--flight dynamic model includes rigid-body modes (plunge, pitch, and their rates) and flexible modes (wing bending and torsion).

The flight conditions for the GLA study are: freestream velocity $U_\infty = 59$~m/s, air density $\rho = 0.0789$~kg/m$^3$, and trim angle of attack $\alpha_0 = 4$~degrees. At these conditions, the aircraft exhibits significant static deformation under the trim loads, with large wing-tip deflections.

\subsection{Open-Loop Stability Analysis}
\label{sec:uav_openloop}

\begin{figure}[htbp]
    \centering
    \includegraphics[width=0.6\textwidth]{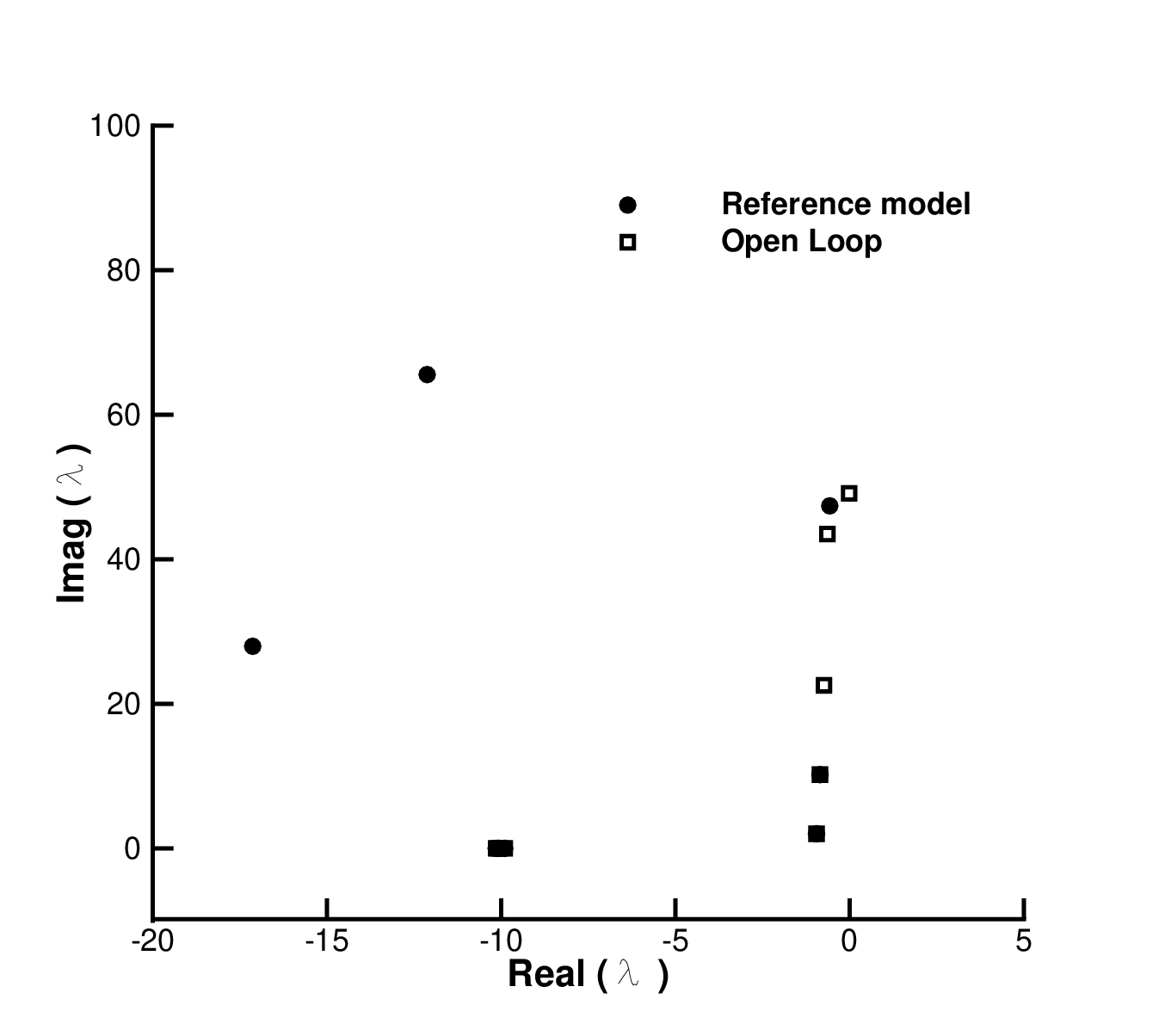}
    \caption{Eigenvalue map of the Global Hawk-like UAV at the trim condition ($U_\infty = 59$~m/s, $\alpha_0 = 4^\circ$). Open-loop eigenvalues (open squares) and reference model eigenvalues (filled circles) are shown. All open-loop eigenvalues have negative real parts, confirming open-loop stability. The reference model eigenvalues are shifted to the left, reflecting the increased damping ratios.}
    \label{fig:openloop_eigs}
\end{figure}

Before designing the adaptive controller, the open-loop stability of the linearised system is assessed. \Cref{fig:openloop_eigs} shows the eigenvalue map of the full-order Jacobian $\mathbf{A}_f$ at the trim condition, with the eight retained ROM eigenvalues highlighted.

All eigenvalues have negative real parts, confirming that the aircraft is open-loop stable at the trim condition. However, several of the aeroelastic modes have low damping ratios, making them susceptible to excitation by gust disturbances. The ROM retains eight physical modes: five oscillatory structural modes (bending and torsion), each represented by a complex-conjugate eigenvalue pair, and three real gust coupling modes near the origin, which together dominate the gust response.

The ROM is validated by comparing its impulse response with the full-order model, showing good agreement for all retained modes. The transmission zeros of the ROM transfer function from flap input to wing-tip displacement output are computed; where non-minimum phase zeros are identified, the Bass--Gura correction described in \cref{sec:mrac_nmp} is applied to ensure the model matching conditions are satisfiable.

\subsection{MRAC Design for the UAV}
\label{sec:uav_mrac_design}

The reference model is designed by increasing the damping ratios of the five oscillatory structural modes (bending and torsion) relative to their open-loop values, while keeping the three real gust coupling modes at their open-loop values. This selective augmentation ensures that the GLA controller targets the structural response without interfering with the gust-induced dynamics.

The matrix $\mathbf{Q}$ is chosen as a diagonal matrix with elements $Q_{ii} = 10^{-4}$, following~\citet{Tantaroudas2017bookchapter}. Three adaptation rates are tested: $\boldsymbol{\Gamma} = 0.01\,\mathbf{Q}$, $0.1\,\mathbf{Q}$, and $1.0\,\mathbf{Q}$.

\subsection{Discrete Gust Response}
\label{sec:uav_discrete}

The discrete gust is a ``1-cosine'' profile defined by:
\begin{equation}
    w_g(t) = \frac{w_{g,\max}}{2}\left(1 - \cos\!\left(\frac{\pi\,U_\infty\,t}{H_g}\right)\right), \qquad 0 \leq t \leq \frac{2\,H_g}{U_\infty}
    \label{eq:1cos_gust}
\end{equation}
where $w_{g,\max}$ is the peak gust velocity and $H_g$ is the gust gradient length. The worst-case gradient length is determined by sweeping $H_g$ over the range specified in the certification requirements and selecting the value that maximises the wing-tip deflection~\citep{DaRonch2013gust}.

\begin{figure}[htbp]
    \centering
    \begin{subfigure}[b]{0.48\textwidth}
        \centering
        \includegraphics[width=\textwidth]{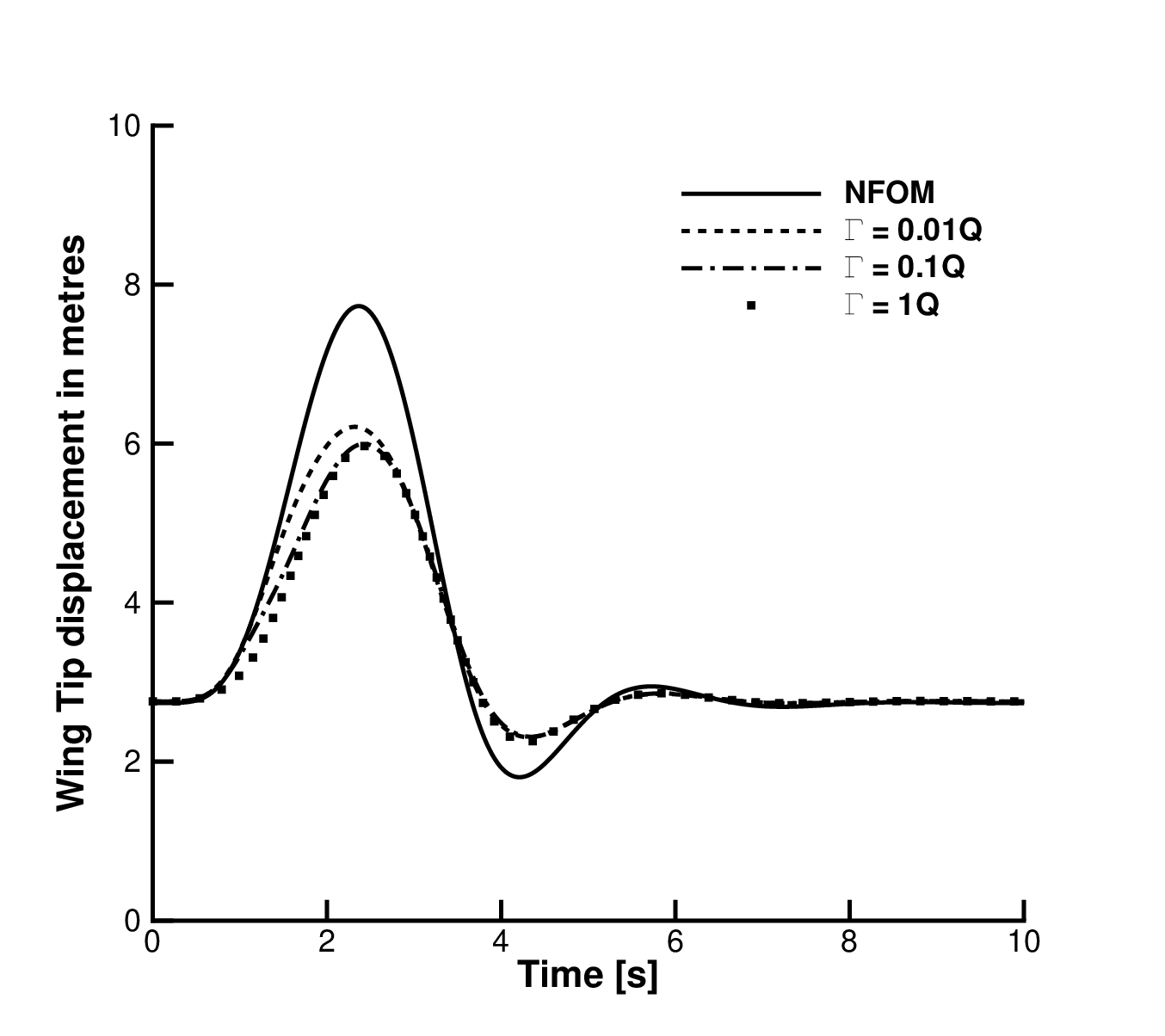}
        \caption{Wing-tip deflection.}
        \label{fig:wtip_mrac_cos}
    \end{subfigure}
    \hfill
    \begin{subfigure}[b]{0.48\textwidth}
        \centering
        \includegraphics[width=\textwidth]{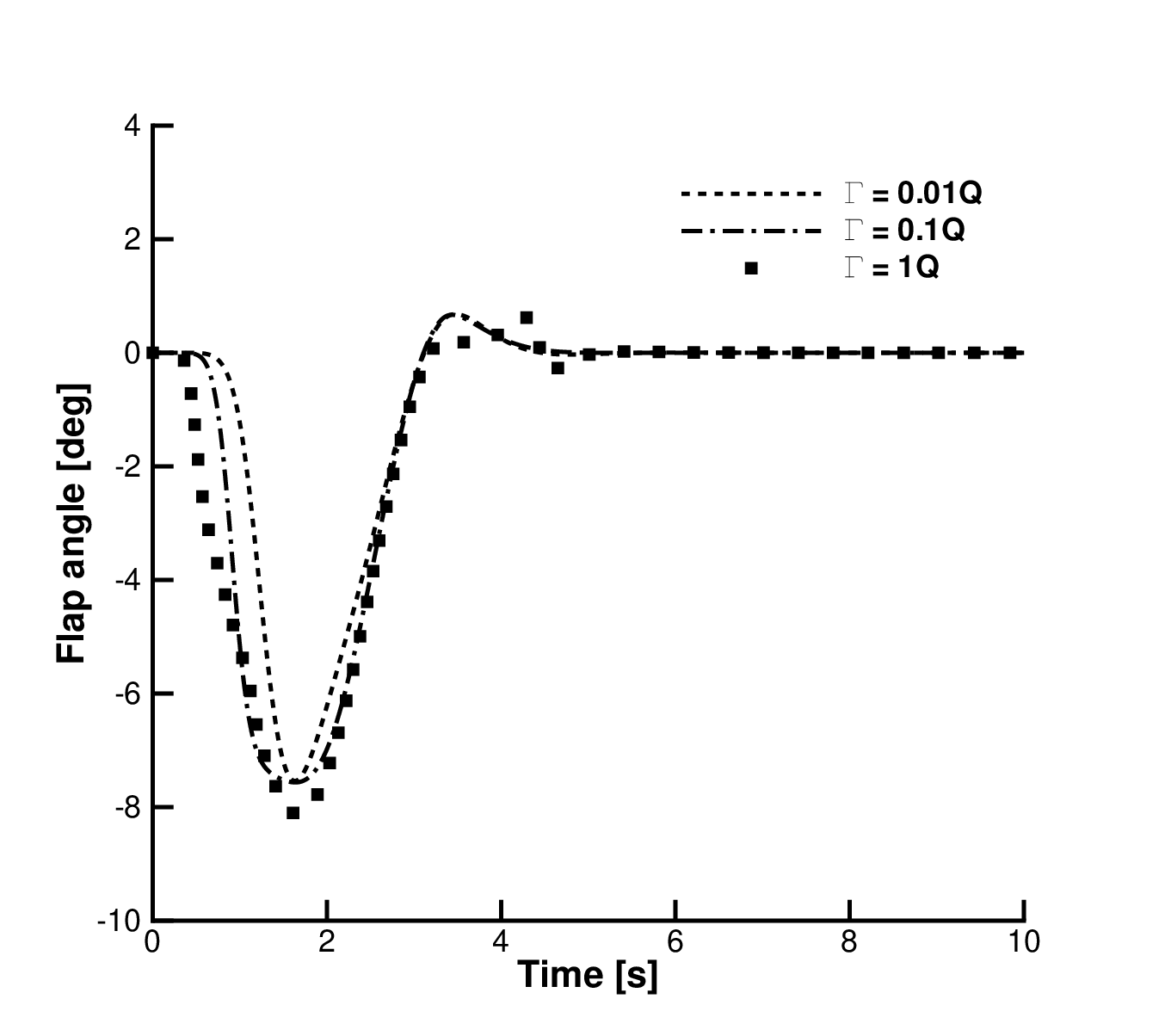}
        \caption{Flap rotation (control input).}
        \label{fig:flap_mrac_cos}
    \end{subfigure}
    \caption{MRAC gust load alleviation for the Global Hawk-like UAV under worst-case discrete (1-cosine) gust: wing-tip deflection and flap rotation for the open-loop system (NFOM) and three adaptation rates ($\boldsymbol{\Gamma} = 0.01\,\mathbf{Q}$, $0.1\,\mathbf{Q}$, $1.0\,\mathbf{Q}$).}
    \label{fig:uav_cos_results}
\end{figure}

\Cref{fig:uav_cos_results} presents the wing-tip deflection and flap rotation time histories under the worst-case discrete gust for the open-loop system and the three MRAC controllers.

\begin{table}[htbp]
    \centering
    \caption{MRAC gust load alleviation performance under worst-case discrete gust for the Global Hawk-like UAV. The wing-tip deflection reduction is measured relative to the open-loop peak.}
    \label{tab:mrac_discrete}
    \begin{tabular}{lcc}
        \toprule
        Controller & Wing-tip reduction (\%) & Max flap (deg) \\
        \midrule
        Open loop & 0 & --- \\
        MRAC, $\boldsymbol{\Gamma} = 0.01\,\mathbf{Q}$ & 24.45 & $-7.54$ \\
        MRAC, $\boldsymbol{\Gamma} = 0.1\,\mathbf{Q}$ & 28.89 & $-7.56$ \\
        MRAC, $\boldsymbol{\Gamma} = 1.0\,\mathbf{Q}$ & \textbf{29.45} & $-8.11$ \\
        $\mathcal{H}_\infty$ robust & 23.15 & $-9.47$ \\
        \bottomrule
    \end{tabular}
\end{table}

\Cref{tab:mrac_discrete} summarises the GLA performance metrics. The highest adaptation rate ($\boldsymbol{\Gamma} = 1.0\,\mathbf{Q}$) achieves the best wing-tip deflection reduction of 29.45\%, with a maximum flap rotation of only $-8.11$ degrees. Even the lowest adaptation rate ($\boldsymbol{\Gamma} = 0.01\,\mathbf{Q}$) provides substantial load alleviation (24.45\%), demonstrating that the MRAC approach is effective across a wide range of $\boldsymbol{\Gamma}$ values. Notably, MRAC with $\boldsymbol{\Gamma} = 1.0\,\mathbf{Q}$ outperforms the $\mathcal{H}_\infty$ robust controller by 6.3 percentage points in wing-tip reduction, while requiring a comparable maximum flap deflection ($-8.11$ vs.\ $-9.47$ degrees). This advantage arises because the adaptive controller can exploit the specific structure of the actual gust encounter, whereas the robust controller must maintain performance for all possible disturbances. The adaptive gains evolve during the gust encounter: they increase as the tracking error grows (when the gust is active) and decay towards their steady-state values as the error diminishes (after the gust has passed). This ``event-driven'' behaviour is a natural consequence of the adaptation law~\eqref{eq:adapt_law}.

\subsection{Stochastic Turbulence Response}
\label{sec:uav_stochastic}

The stochastic turbulence model uses a Von K\'{a}rm\'{a}n power spectral density with parameters appropriate for the flight altitude~\citep{Tantaroudas2014aviation}. A realisation of the turbulence time history is generated and applied to the aircraft model.

\begin{figure}[htbp]
    \centering
    \begin{subfigure}[b]{0.48\textwidth}
        \centering
        \includegraphics[width=\textwidth]{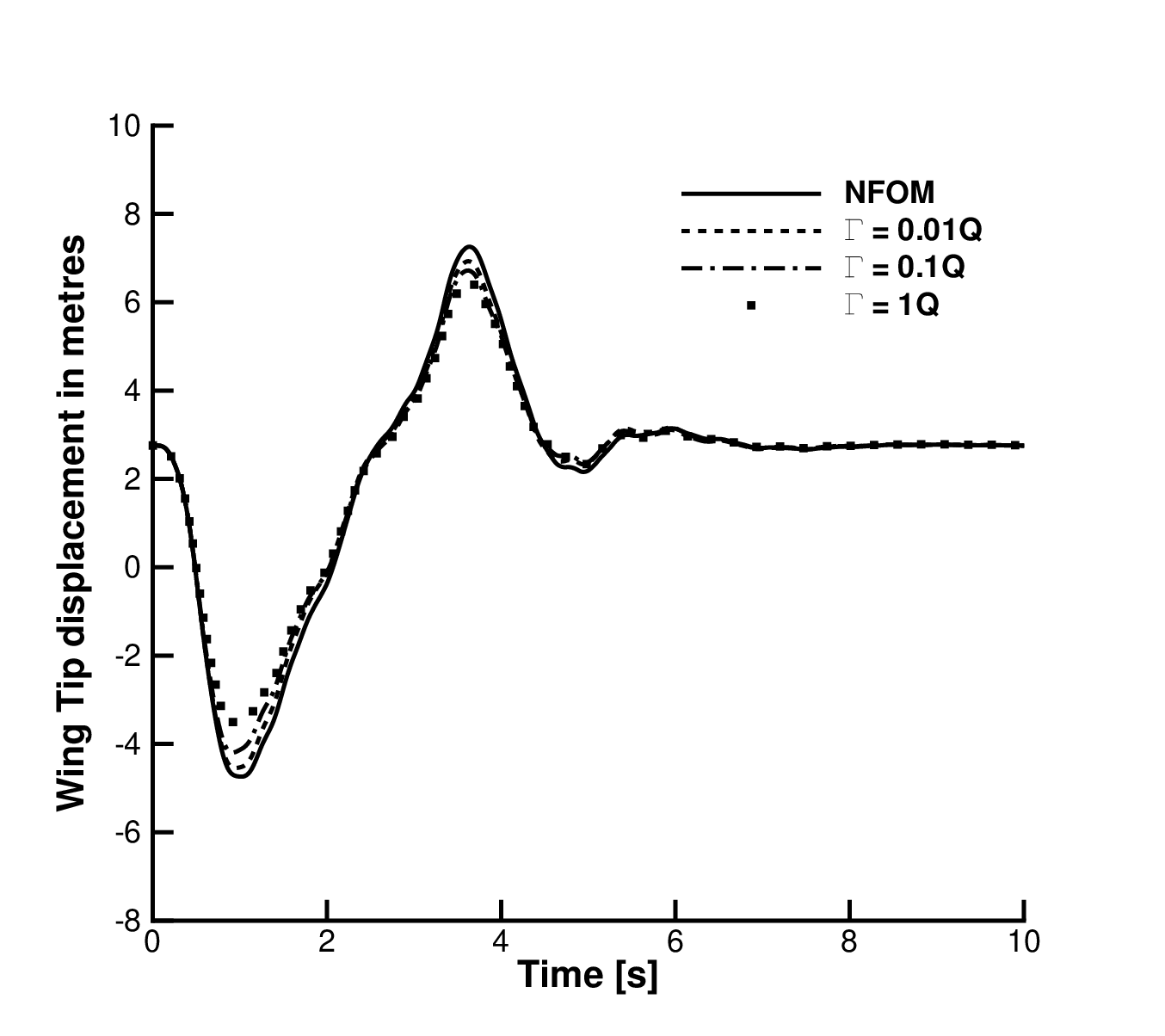}
        \caption{Wing-tip deflection.}
        \label{fig:wtip_mrac_vk}
    \end{subfigure}
    \hfill
    \begin{subfigure}[b]{0.48\textwidth}
        \centering
        \includegraphics[width=\textwidth]{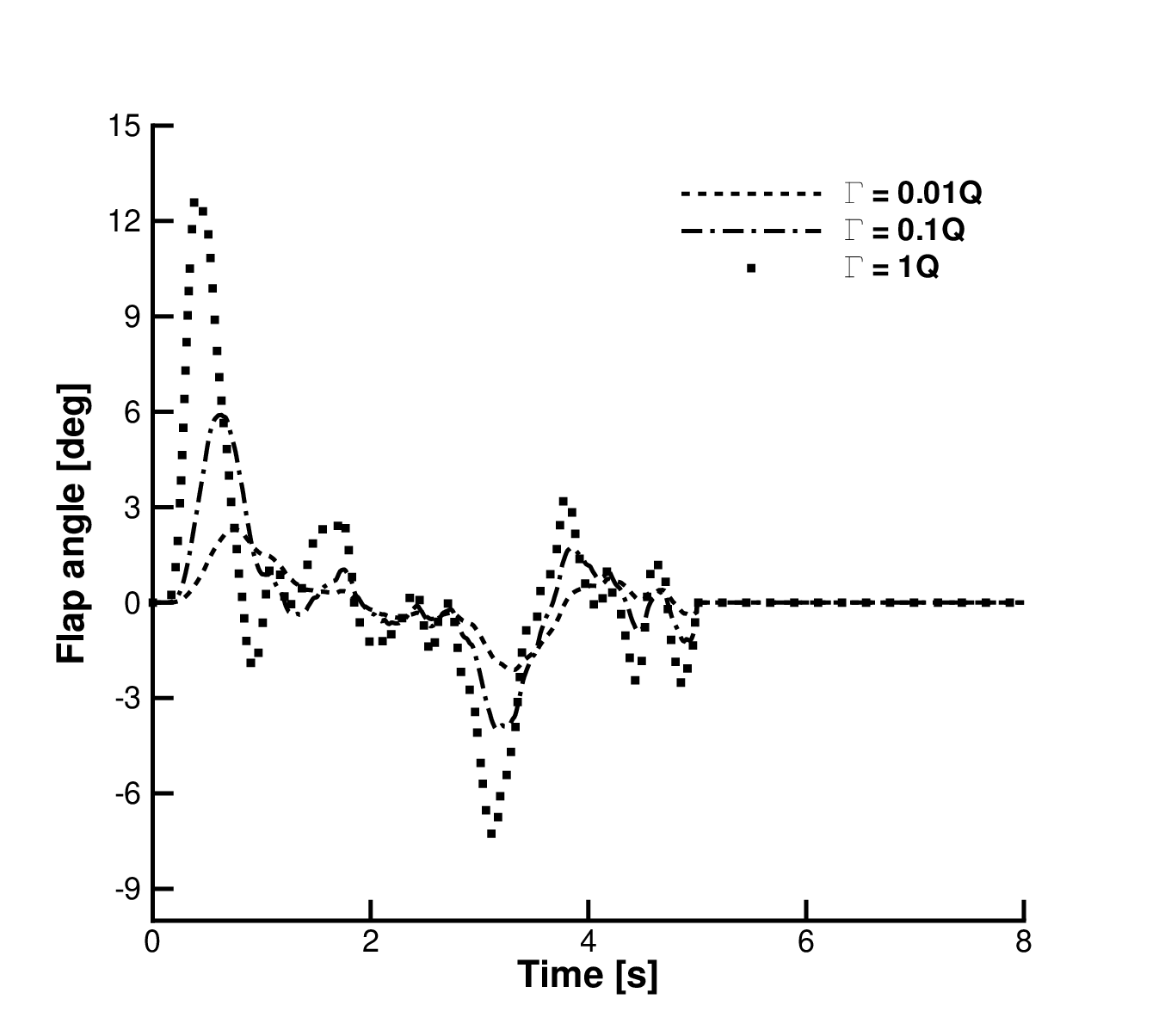}
        \caption{Flap rotation (control input).}
        \label{fig:flap_mrac_vk}
    \end{subfigure}
    \caption{MRAC gust load alleviation for the Global Hawk-like UAV under Von K\'{a}rm\'{a}n stochastic turbulence: wing-tip deflection and flap rotation for the open-loop system (NFOM) and three adaptation rates. The initial transient response is shown.}
    \label{fig:uav_vk_results}
\end{figure}

\Cref{fig:uav_vk_results} presents the wing-tip deflection and flap rotation time histories under the stochastic turbulence for the open-loop system and the three MRAC controllers.

\begin{table}[htbp]
    \centering
    \caption{MRAC gust load alleviation performance under Von K\'{a}rm\'{a}n stochastic turbulence for the Global Hawk-like UAV. The wing-tip deflection reduction is measured relative to the open-loop value.}
    \label{tab:mrac_stochastic}
    \begin{tabular}{lcc}
        \toprule
        Controller & Wing-tip deflection reduction (\%) & Max flap (deg) \\
        \midrule
        MRAC, $\boldsymbol{\Gamma} = 0.01\,\mathbf{Q}$ & 4.73 & 2.31 \\
        MRAC, $\boldsymbol{\Gamma} = 0.1\,\mathbf{Q}$ & 8.00 & 5.89 \\
        MRAC, $\boldsymbol{\Gamma} = 1.0\,\mathbf{Q}$ & \textbf{12.68} & 12.83 \\
        $\mathcal{H}_\infty$ robust & 10.26 & 12.79 \\
        \bottomrule
    \end{tabular}
\end{table}

The stochastic turbulence results (\cref{tab:mrac_stochastic}) reveal several important differences from the discrete gust case. The overall load reductions are smaller than for the discrete gust (12.68\% vs.\ 29.45\% at the highest adaptation rate), which is expected because the stochastic turbulence contains energy across a broad frequency band, and the adaptive controller cannot fully anticipate the rapidly varying disturbance. The low adaptation rate ($\boldsymbol{\Gamma} = 0.01\,\mathbf{Q}$) provides only 4.73\% reduction, which is insufficient for practical GLA, since the gains simply cannot adapt quickly enough to track the rapidly changing gust environment; this highlights the importance of selecting an appropriate adaptation rate for the expected disturbance spectrum. The $\mathcal{H}_\infty$ controller achieves 10.26\% wing-tip reduction with a maximum flap rotation of 12.79 degrees: for larger adaptation rates ($\boldsymbol{\Gamma} = 1.0\,\mathbf{Q}$), the MRAC achieves a comparable level of gust load alleviation (12.68\%) with a similar control effort (12.83 degrees), while for lower adaptation rates the MRAC performance degrades below that of $\mathcal{H}_\infty$. The high adaptation rate ($\boldsymbol{\Gamma} = 1.0\,\mathbf{Q}$) requires continuous, large-amplitude flap activity, which may raise concerns about actuator fatigue and energy consumption in prolonged turbulence encounters, while the moderate rate ($\boldsymbol{\Gamma} = 0.1\,\mathbf{Q}$) offers a better trade-off, providing 8.00\% reduction with substantially smaller flap demands. Under stochastic turbulence, the adaptive gains exhibit continuous variation, unlike the transient-then-decay behaviour observed for the discrete gust. The gains track the turbulence-induced tracking error, effectively providing a time-varying extension of the fixed reference model, which is both a strength (enabling real-time adjustment) and a potential limitation (sensitivity to noise).

%% ============================================================
\section{Discussion}
\label{sec:discussion}
%% ============================================================

\subsection{Adaptation Rate Selection Guidelines}
\label{sec:discussion_gamma}

The results from both test cases establish guidelines for adaptation rate selection in MRAC for GLA. For discrete (deterministic) gusts, the adaptation rate can be selected based on the expected worst-case gust: higher rates provide better performance, and the transient nature of the disturbance limits the total control effort. For the UAV test case, $\boldsymbol{\Gamma} = 1.0\,\mathbf{Q}$ provides the best performance with acceptable actuator demands, while increasing $\boldsymbol{\Gamma}$ beyond this value yields only marginal additional gains with substantially increased flap demands. For stochastic turbulence, the continuous nature of the disturbance makes adaptation rate selection more critical, as high rates provide the best instantaneous load reduction but at the cost of continuous, large-amplitude control activity; for sustained turbulence encounters, $\boldsymbol{\Gamma} = 0.1\,\mathbf{Q}$ offers a practical compromise. In operational service, aircraft encounter both discrete gusts and continuous turbulence, suggesting that a gain-scheduling approach, in which $\boldsymbol{\Gamma}$ is increased when a discrete gust is detected and reduced during continuous turbulence, would combine the benefits of both regimes.

\subsection{Comparison with \texorpdfstring{$\mathcal{H}_\infty$}{H-infinity} Robust Control}
\label{sec:discussion_hinf}

The comparison with $\mathcal{H}_\infty$ robust control (\cref{tab:mrac_discrete}) demonstrates a clear advantage for MRAC under discrete gusts: 29.45\% vs.\ 23.15\% wing-tip reduction with comparable control effort. This advantage arises from two sources. First, the MRAC gains adjust in real time to the specific gust encounter, whereas the $\mathcal{H}_\infty$ controller uses fixed gains designed for worst-case performance across all anticipated disturbances, and this conservatism limits the achievable performance of the robust controller. Second, the MRAC reference model explicitly specifies the desired gust response through the increased damping ratios, whereas the $\mathcal{H}_\infty$ controller optimises a weighted combination of performance and robustness objectives that may not directly correspond to the desired GLA behaviour.

However, the $\mathcal{H}_\infty$ controller has the advantage of guaranteed stability margins and does not require online gain adaptation, making it simpler to certify. In practice, a combined approach, using $\mathcal{H}_\infty$ as a baseline controller with MRAC providing augmentation during severe gust encounters, may be the most practical solution~\citep{Tantaroudas2015scitech, Tantaroudas2017bookchapter}.

\subsection{Effect of Nonlinearities}
\label{sec:discussion_nonlinear}

The cubic hardening nonlinearities in the 3-DOF aerofoil test case have a stabilising effect at large deformations, effectively increasing the structural stiffness and limiting the amplitude of the gust-induced oscillations. The MRAC controller operates effectively in the presence of these nonlinearities, with the differential nonlinearity term $\mathbf{F}_{Df}$ remaining within the Lipschitz bound~\eqref{eq:lipschitz} throughout all simulations.

For the UAV test case, the geometric nonlinearities arising from large wing deformations are milder than the cubic terms in the aerofoil model, and the linearised ROM provides a good approximation of the nonlinear dynamics in the neighbourhood of the trim condition. The MRAC stability condition~\eqref{eq:lipschitz} is satisfied with a comfortable margin for all gust amplitudes considered~\citep{Tantaroudas2014aviation}.

\subsection{Scalability and Practical Considerations}
\label{sec:discussion_practical}

The MRAC framework presented in this work scales well to higher-dimensional systems. The computational cost is dominated by the ROM size $n$ (typically 4--8 for the applications considered), not by the full-order model size $N$. The main practical challenges for implementation on a real aircraft are threefold. First, full-state feedback is assumed in the current formulation, whereas in practice the ROM states must be estimated from sensor measurements using an observer such as a Kalman filter; the observer design and its interaction with the adaptive law is an important consideration beyond the scope of this paper. Second, the current formulation assumes ideal actuators with no rate or magnitude limits, whereas in practice actuator dynamics and constraints must be incorporated, potentially through anti-windup modifications to the adaptation law. Third, the MRAC formulation requires that the model matching conditions~\eqref{eq:matching_Kx}--\eqref{eq:matching_Kr} are satisfiable; if the plant model changes significantly (e.g., due to damage or configuration change), the ROM may need to be updated, and the adaptation law will need to compensate for the model mismatch.

%% ============================================================
\section{Conclusions}
\label{sec:conclusions}
%% ============================================================

Model Reference Adaptive Control based on Lyapunov stability theory has been developed and validated for gust load alleviation of nonlinear aeroelastic systems. The complete MRAC formulation, including reference model design, adaptive control law, error dynamics, Lyapunov stability proof, and adaptation law derivation, has been presented for coupled aeroelastic--flight dynamic systems with nonlinear reduced-order models.

The key findings and contributions are as follows. For the Global Hawk-like UAV, MRAC with $\boldsymbol{\Gamma} = 1.0\,\mathbf{Q}$ achieves 29.45\% wing-tip deflection reduction under the worst-case discrete gust, outperforming $\mathcal{H}_\infty$ robust control (23.15\%) with comparable control effort, thereby demonstrating the advantage of adaptive over fixed-gain control for deterministic disturbances. Under Von K\'{a}rm\'{a}n stochastic turbulence, MRAC with $\boldsymbol{\Gamma} = 1.0\,\mathbf{Q}$ achieves 12.68\% wing-tip deflection reduction, with performance decreasing to 4.73\% at $\boldsymbol{\Gamma} = 0.01\,\mathbf{Q}$; the trade-off between load reduction and actuator demand is governed directly by the adaptation rate. The adaptation rate $\boldsymbol{\Gamma}$ emerges as the critical design parameter, governing the convergence speed, peak load reduction, and actuator demands: for discrete gusts, $\boldsymbol{\Gamma} = 1.0\,\mathbf{Q}$ is recommended, while for sustained turbulence, $\boldsymbol{\Gamma} = 0.1\,\mathbf{Q}$ provides a practical compromise. The stability proof for the nonlinear case establishes that the tracking error remains bounded provided the differential nonlinearity satisfies $\|\mathbf{F}_{NR}(\mathbf{x}) - \mathbf{F}_{NR}(\mathbf{x}_m)\| \leq L_F\,\|\mathbf{x} - \mathbf{x}_m\|$, with the Lipschitz constant $L_F = \lambda_{\min}(\mathbf{Q})/(2\|\mathbf{P}\|)$ determined by the reference model design; this condition is satisfied for typical aeroelastic nonlinearities in a neighbourhood of the equilibrium. The framework is general and computationally efficient, operating on a small reduced-order model ($n = 8$--$16$ states) suitable for real-time implementation and applicable to any coupled aeroelastic--flight dynamic system expressible in state-space form.

Future work will address gain-scheduled adaptation rates for mixed discrete/stochastic environments, observer-based implementation with output feedback, and experimental validation on a wind tunnel aeroelastic model.

\section*{Acknowledgements}

This work was supported by EPSRC grant EP/I014594/1 (Nonlinear Flexibility Effects on Flight Dynamics and Control of Next-Generation Aircraft). The authors are grateful to Prof.\ A.\ Da Ronch and Prof.\ K.J.\ Badcock for their valuable guidance on flight dynamics modelling.

\bibliographystyle{unsrtnat}
\bibliography{references}

\end{document}